\documentclass[twoside,twocolumn,english,prl,superscriptaddress,nobibnotes,nofootinbib,floatfix]{revtex4-2}
\usepackage{mathtools}
\usepackage{amsmath}
\usepackage{amssymb}
\usepackage{enumitem}
\usepackage{dsfont}
\usepackage[usenames,dvipsnames]{color}
\usepackage{pythonhighlight}
\usepackage[pdftex,unicode=true, bookmarks=false, 
breaklinks=false,pdfborder={0 0 0}, colorlinks=false] {hyperref}
\usepackage[caption=false]{subfig}
 \usepackage{graphicx}
\usepackage{newtxtext,newtxmath}
\usepackage{scalerel}
\usepackage{stackengine}
\usepackage{braket}

\usepackage{epigraph}
\usepackage{babel}

\newcommand{\smallselfleft}{\rotatebox[origin=c]{270}{{\scaleobj{0.6}{\circlearrowright}}}}
\newcommand{\smallselfright}{\rotatebox[origin=c]{90}{{\scaleobj{0.6}{\circlearrowleft}}}}

\def\hri#1#2{\href{http://arxiv.org/abs/#1}{[ArXiv:#1]#2}}
\def\hre#1#2{\href{http://arxiv.org/abs/#1/#2}{[ArXiv:#1/#2]}}

\hypersetup{
 pdfauthor={Leandro Silva Pimenta},
 pdfsubject={Physics},
 pdfkeywords={quantum, stochastic, indivisible, dynamics, causality, continuity, cone}}

\begin{document}
\title{Divisible and indivisible Stochastic-Quantum dynamics}
\author{Leandro Silva Pimenta}
\email{leandro.silva-pimenta@espci.org}
\affiliation{Independent researcher}

\begin{abstract}
This work presents a complete geometrical characterisation of divisible and indivisible time-evolution at the level of probabilities for systems with two configurations, open or closed. Our new geometrical construction in the space of stochastic matrices shows the existence of conical bounds separating divisible and indivisible dynamics, bearing analogy with the relativistic causal structure, with an emerging time pointing towards information erasure when the dynamics are divisible. Indivisible dynamics, which include quantum dynamics, are characterised by a time-flow against the information-erasure time coordinate or by being tachyonic with respect to the cones in the stochastic matrix space. This provides a geometric counterpart of other results in the literature, such as the equivalence between information-decreasing and divisible processes. The results apply under minimal assumptions: (i) the system has two configurations, (ii) one can freely ascribe initial probabilities to both and (iii) probabilities at other times are linearly related to the initial ones through conditional probabilities. The optional assumption of (iv) continuity places further constraints on the system, removing one of the past cones. Discontinuous stochastic dynamics in continuous time include cases with divisible blocks of evolution which are not themselves divisible. We show that the connection between continuity and multiplicity of divisors holds for any dimension. We extend methods of coarse graining and dilations by incorporating dynamics and uncertainty, connecting them with divisibility criteria. This is a first step towards a full geometric characterisation of indivisible stochastic dynamics for any number of configurations which, as they cannot at the level of probabilities be reduced to a composition of evolution operators, constitute fundamental elements of probabilistic time-evolution.
\end{abstract}
\maketitle

\setlength{\epigraphwidth}{.41\textwidth}

\epigraph{We call continuous that whose change in its own right is one and cannot be otherwise; and a change is one when indivisible, and indivisible in respect of time. [...] Now what is indivisible in respect of quantity in all dimensions is called a unit if it has no position, a point if it has position, and what is divisible with respect of quantity in one dimension is a line, in two a plane, in all three a body.
}{Aristotle, Metaphysics, book $\Delta$, 6, 1016b \protect\cite{Aristotle}}

\section{Introduction\label{sec:Introduction}}

For over a century of applications of stochastic matrices since Markov's 1906 seminal work \cite{Markov:1906, Seneta:1996}, probabilities in physics are split: celebrated in the classical realm and relegated to a secondary rank at the quantum level where they are obtained from amplitudes via the Born rule.
The origin of this incompatibility which prevented dealing, for instance, with interference from a purely probabilistic point of view, was a hypothesis that, once dropped, allows for a derivation of the Quantum formalism from probability theory \cite{Barandes:2023ivl,Barandes:2023pwy}. The assumption to be abandoned is the {\it divisibility} of the stochastic dynamical map. Indivisibility was first identified in the study of quantum channels \cite{Wolf:2008kbg} and has been an active area of research \cite{Plenio:2014nyj,Breuer:2015,Buscemi:2016,Davalos:2018,LiLi:2018ysq, Chruscinski:2022hvy,Ende:2024ksq, Nery:2024fyl}. In this work we present new results regarding the divisibility of the stochastic dynamics for virtually any system that can be described by two configurations and admit a probabilistic description, and argue in favour of the stochastic point of view as providing a unifying, model-independent approach that may lead to insights connecting different areas of physics.

It comes with great surprise that, despite a century of Markovian dynamics, the study of indivisible stochastic dynamics is at its infancy. With the exception of $2\times 2$ and $3\times 3$ matrices \cite{vomEnde:2024isq}, and few special cases in every dimension\footnote{Every permutation matrix is both column and row-stochastic. The group structure guarantees that all $N\times N$ permutation matrices are divisible stochastic matrices. They are also unitary and a play a role in the divisibility of stochastic matrices analogous to the one of unitary matrices for quantum channels \cite{Davalos:2018} which we develop in section \ref{ssec:symmetries}.}, the explicit form of divisible stochastic matrices is unknown. Other general results, such as the equivalence between divisibility of a stochastic matrix and information reduction \cite{Buscemi:2016} may provide a test for the divisibility of a process. However, deriving the general form of the indivisible matrices from this criterion is yet to be done. A step further consists in characterising an indivisible stochastic {\it evolution} in the space of stochastic matrices as it requires further constraints. Here we show that divisibility entails the existence of a cone structure in the space of $2\times2$ stochastic matrices, with past and future cones, and regions between them that are analogous to space-like separation in relativity. The information-theoretical future points towards erasure of the initial information, with indivisible dynamics displaying either an opposition between the time-parameter of the evolution and the information-erasure time {\it or} a ``tachyonic" behaviour in matrix space\footnote{This is {\it not} a space-time description but a description in the space of matrices of conditional probabilities or, equivalently, stochastic matrices. Nothing prevents to apply our results in space-time through, for instance, the bipartition of a region (see \cite{Plenio:2001,Kastner:2023} in relation with the irreversible cases of section \ref{sec:two_configurations} here) as long as the answer to the question asked is binary and satisfies the other minimal assumptions.}. This shows how an information-theoretical notion bears an unsuspected analogy with relativistic causality when one considers (in)divisibility from a geometrical point of view.

Accepting indivisible stochastic processes opens the door for a richer phenomenology that incorporates quantum phenomena and non-Markovian systems\footnote{The quantum formalism can be seen as the analytical mechanics of indivisible stochastic processes from the perspective of the Stochastic-Quantum Correspondence \cite{Barandes:2023ivl,Barandes:2023pwy}. This dictionary opens up new perspectives for the study of systems with memory in all fields, as indivisibility is a necessary condition for non-Markovianity even at the quantum level where there has an interesting discussion about the meaning non-Markovianity \cite{Plenio:2014nyj,Breuer:2015,Buscemi:2016,LiLi:2018ysq, Chruscinski:2022hvy}.} which are the rule, while Makovian ones are the exception \cite{vanKampen:1998}.
This is a first step towards the general geometric characterisation of the fundamental blocks of indivisible stochastic dynamics, the atoms of probabilistic evolution. As such, the present work is inscribed in a chain of efforts, dating back at least to Aristotle \cite{Aristotle}, aiming at identifying fundamental and indivisible elements of our world.

Given initial standalone probabilities $p_j(0)$ and evolved probabilities $p_i(t)$, where $i$ and $j$ range over the configurations $1, 2, \dots, N$, one can relate them through Bayesian marginalisation.
\begin{equation}
p_{i}\left(t\right)=\sum_{j=1}^{N}p\left(i, t | j, 0\right)p_{j}\left(0\right)
\label{p_evolution}
\end{equation}
{\bf Divisibility} of the stochastic dynamics at time $t^\prime$ means that the matrix of conditional probabilities $p\left(i, t | j, 0\right)$ can be split at this specific instant.
\begin{align}
&p\left(i, t | j, 0\right) \text{~divisible at~}t=t^\prime
\quad\iff
\nonumber \\
&p\left(i, t | j, 0\right) =\sum_{k=1}^N p\left(i, t | k, t^\prime \right)p\left(k, t^\prime | j, 0\right), \quad 0\leqslant t^\prime \leqslant t
\label{divisibility_probabilities}
\end{align}
When \eqref{divisibility_probabilities} holds, $t^\prime$ is called a {\bf division event}\footnote{A relativistically inclined reader may be concerned about the notion of an event that extends over configurations, as we would be using a preferred foliation in a continuum limit. At this point there are no known objections to consider different foliations in this setup nor it is understood how stochastic divisibility for a continuous system changes with a foliation change. It is not, however, unphysical to consider a preferred foliation as, for instance, the CMB is not isotropic for every family of observers. Finally, since leaves of constant mean-curvature, the so called CMC gauge, are singled out when considering the configuration space of General Relativity as the conformal superspace \cite{Barbour:2010xk}, this setup in a continuum limit may be sufficient for extending indivisible stochastic dynamics to the gravitational realm.}, following Barandes \cite{Barandes:2023ivl,Barandes:2023pwy}; when it does not, we say the evolution is {\bf indivisible} at time $t^\prime$. This definition referring to {\it specific instants} and not to the entire evolution of the system does not prevent one for requiring it to hold for all times, further restricting the space of processes in question, in which case we will talk about a divisible {\it process}, following Plenio et al. \cite{Plenio:2014nyj}.

With the validity or failure of \eqref{divisibility_probabilities} evaluated at each instant, a generic stochastic evolution will involve periods of indivisible evolution separated by periods of divisible evolution. In this sense, what \cite{Plenio:2014nyj} calls an indivisible process is, generically, a combination of purely divisible and purely indivisible periods of stochastic evolution.

Given a quantum system, open or closed, and a dynamical map describing its evolution, it is trivial to obtain from it stochastic dynamics by fixing a basis. More precisely, given any positive trace-preserving map $\Phi$ from linear operators to linear operators in a Hilbert space, and a basis $\set{\ket{k}}_{k=1}^N$, there is a left-stochastic matrix $\Gamma$ such that
\begin{equation}
\Gamma_{ij} = \text{tr}\left[\ket{i}\bra{i}\Phi\left(\ket{j}\bra{j}\right)\right]
\label{channel_to_Gamma}
\end{equation}
with different matrices for different basis \cite{Chruscinski:2022hvy}. One may also perform this transformation directly on a given Kraus representation of a quantum channel by intercalating it with a pair of L\"uders operations \cite{Schmidt:2021}.

Going from stochastic to quantum dynamics is also possible. In \cite{Barandes:2023ivl,Barandes:2023pwy} the author presents a prescription to build a quantum dynamical map from indivisible stochastic dynamics and, using Stinespring's dilation theorem \cite{Stinespring:1955}, build a Hilbert space formulation with a unitary time-evolution. This connection between two apparently distinct kinds of dynamics is called the Stochastic-Quantum Correspondence (SQC).

In \cite{Barandes:2023ivl,Barandes:2023pwy} the author associates between quantum phenomena such as interference with a manifestation of indivisibility of the time evolution as in equation \eqref{divisibility_probabilities}. In subsection \ref{sec:consequences_stochastic-quantum} we apply our results to show that this association is more delicate, since division events may occur for time intervals during which the density matrix has non-zero coherences according to the very prescription of \cite{Barandes:2023ivl,Barandes:2023pwy}. Our analysis also shows, and we develop it in subsection \ref{ssec:maximal_maps} a geometric constraint on the allowed division events for \eqref{divisibility_probabilities} to hold for every $t\geqslant t^\prime$. In \cite{Barandes:2024ddf} the time non-locality which is inherent to indivisible dynamics is used to show how indivisible stochastic dynamics evade Bell's theorems \cite{Bell:1964kc,Bell:1975uz}. Since we show here that not all division events are equal, it would be interesting to understand how this analysis depends on the level of indivisibility of a Bell non-local system and its subsystems.

Working solely with probabilities at the level of the stochastic dynamics has the advantage of providing results which are, therefore, valid for classical, semi-classical and quantum systems. For these reasons, the remainder of this paper will deal with divisibility or indivisibility at the level of the {\it stochastic} dynamics, except in section \ref{sec:consequences_stochastic-quantum} where we discuss connections with and implications for the SQC. For a quantum-informational perspective with emphasis on the divisibility of quantum channels and dynamical maps, see \cite{Wolf:2008kbg,Plenio:2014nyj,LiLi:2018ysq, Chruscinski:2022hvy,Ende:2024ksq, Nery:2024fyl} and references therein.

The present work also extends the results of \cite{vomEnde:2024isq} by incorporating dynamics and showing that despite the fact that all $2\times2$ stochastic matrices are divisible, this is not the case for the stochastic time-evolution and there are almost always\footnote{The space of matrices being represented as a square, the exceptions are located on a line. If we use the Euclidean area as a measure over the space of matrices, the set of exceptions has therefore measure zero. They play nevertheless a capital role that we shall see in section \ref{sec:two_configurations}.} two disjoint regions in the space of $2\times2$ stochastic matrices that contain divisible dynamics. We also show that these are the intersections of the space of stochastic matrices with cones generated by rays emanating from the irreversible deterministic cases and passing through the present value of the stochastic evolution matrix, plus a second region obtained by permutation. This geometric picture allows for a clear understanding of the consequences of continuity.

With the problem for two configurations fully solved, this work deals with larger systems by coarse graining and dilation, extending the approach of Schmidt \cite{Schmidt:2021} of building environment dilations by reversing a coarse-graining. The first generalisation involves dynamics and we show that the degenerate cases for $2\times2$ stochastic matrices all arise as dynamical dilations. With this done, we present a set of equations relating divisibility, dilations and coarse graining. For the latter, we present a  second generalisation of \cite{Schmidt:2021} which consists in the inclusion of uncertainty in the coarse-graining procedure which serves to identify the structure of some solutions to the divisibility equation and build from it divisible dynamics.

This paper is organised as follows. Section \ref{sec:divisible_indivisible} presents a definition of divisible and indivisible stochastic dynamics, introduces a few elements and reviews some results in the literature regarding uses and practical characterisations of indivisibility. Section \ref{sec:two_configurations} contains the main results of this paper. It starts with a characterisations of stochastic systems with $2$ configurations from a physical point of view, associating stochastic matrices and graphs. The familiar reader may go directly to subsection \ref{ssec:divisible_two} where we present a new geometric characterisation of necessary and sufficient conditions for indivisibility of the stochastic dynamics for systems with two configurations. In subsection \ref{ssec:continuous_two} we connect these results with the continuity and the constraints they imply for time-evolution. In subsection \ref{ssec:density_connectivity} we consider the connectivity of the possible divisors of divisible matrix and their relative densities through examples. We provide symmetry considerations valid for any (finite) dimension in subsection \ref{ssec:symmetries} and apply them to partly explain the structure of the connections shown in the preceding subsection. In section \ref{sec:information} we show how the geometric construction naturally incorporates information reduction and we rewrite the matrices using an ``information-reduction" time coordinate and an asymmetry parameter which plays the role of a spatial dimension, connecting some results with information inequalities and the graph decomposition of section  \ref{sec:divisible_indivisible}. Other information inequalities appear in section \ref{sec:consequences_stochastic-quantum} where we consider the consequences of the present analysis to the Stochastic-Quantum Correspondence.  We discuss coarse graining of larger systems, dilation by coarse graining and their connection with divisibility in section \ref{sec:coarse} where we apply and extend the results of \cite{Schmidt:2021}. Section \ref{sec:conclusions_and_outlook} concludes the paper and provides perspectives for future research.

\section{Divisible and indivisible dynamics}
\label{sec:divisible_indivisible}

Here we review the notion of a generalised stochastic system, of indivisible time-evolution, summarising main results from the literature, and introduce some elements of the Stochastic-Quantum Correspondence (SQC) that allow these results to have wide range of applications.

\subsubsection{Generalised stochastic systems}

Physical systems with ascribed probabilistic evolutions are omnipresent, either by a measure of our ignorance being expressed through the equations, or because of intrinsic stochasticity influencing their evolution. In this section we present a summary of Barandes' definition of a generalised stochastic system \cite{Barandes:2023ivl}.

Consider a system with $N$ physically distinct configurations such that one may ascribe standalone probabilities to each one of them at an initial time $t=0$ which may correspond, for instance, to a preparation.

The standalone probability of configuration $i$ at time $t$ is denoted by $p_{i}(t)$.
\begin{equation}
p_{i}(t)\geqslant 0, \qquad \sum_{j=1}^N p_{i}(t)= 1
\end{equation}
The $p_{i}(t)$ are linearly related to the initial ones, at time $t=0$, by Bayesian marginalisation, equation \eqref{p_evolution}, implying a division event at $t=0$. Following Barandes \cite{Barandes:2023ivl}, hereafter we denote the matrix of conditional probabilities in \eqref{p_evolution} by
\begin{equation}
\Gamma_{ij}(t)\equiv p\left(i,t | j,0\right)
,\label{gamma_ij_p_j}
\end{equation}
and rewrite \eqref{p_evolution} as an equation of matrix multiplication.
\begin{equation}
p\left(t\right)=\Gamma\left(t\right)p\left(0\right)
\label{Gamma_p_compact}
\end{equation}

If one does not know about \eqref{p_evolution} but requires a linear evolution of the probabilities according to \eqref{Gamma_p_compact}, the possibility that the initial system is with certainty in any of the initial configurations, forces any map $\Gamma$ between probabilities of the form \eqref{Gamma_p_compact} to have columns that are, themselves, positive entries adding to one.
\begin{equation}
\Gamma_{ij}\geq0,\qquad\sum_{i=1}^{N}\Gamma_{ij}=1
\label{left-stochastic}
\end{equation}
Conditions \eqref{left-stochastic} define $\Gamma$ as a {\it left-stochastic matrix}, also known as {\it column-stochastic}. If, instead of acting on probability column vectors from the left, as in \eqref{Gamma_p_compact}, $\Gamma$ acted on probability row vectors from the right, the resulting equation would be simply the transpose of \eqref{Gamma_p_compact}, thus defining transformations with non-negative entries and rows with unit sum, the so-called {\it right-stochastic matrix}. A matrix that is both left and right stochastic is called {\it doubly stochastic} or {\it bistochastic}.

A {\bf generalised stochastic system} is a tuple $(\mathcal{C}, \mathcal{T}, \Gamma, p,\mathcal{A})$ such that $\mathcal{C}$ is the collection of the system's allowed configurations, $\mathcal{T}$ is the collection of times for which the system is defined, $\Gamma$ are the stochastic maps containing the conditional probabilities as a function of time (hence a map from $\mathcal{C}^2\times\mathcal{T}\to [0,1]$) defined in \eqref{gamma_ij_p_j} or by \eqref{left-stochastic}, $p$ denotes the standalone probabilities $p_i(t)$ of which only $p_i(0)$ are supposed to be freely adjustable, related to each other via \eqref{Gamma_p_compact}, and $\mathcal{A}$ is the system's algebra of random variables \cite{Barandes:2023pwy}. we will not make use of $\mathcal{A}$ in this paper and the collection $\mathcal{T}$ being used will be specified when needed.

\subsection{Indivisible matrices vs. indivisible dynamics}
\label{ssec:divisible_and_indivisible}

An $N\times N$ stochastic {\it matrix} $\Gamma$ is defined by \eqref{left-stochastic} and is called {\it divisibile} if there are $N\times N$ stochastic matrices $\hat\Gamma$ and $\tilde\Gamma$ such that
\begin{equation}
\Gamma = \hat\Gamma~ \tilde\Gamma.
\label{matrix_divisibility}
\end{equation}
Every $2\times 2$ stochastic matrix is divisible \cite{vomEnde:2024isq}.

A stochastic dynamics given by a collection of stochastic matrices $\Gamma(t)$ parametrised by time $t$ is divisible at time $t^\prime$ if there is stochastic matrix $\Gamma(t \leftarrow t^\prime)$ such that
\begin{equation}
\Gamma(t) = \Gamma(t \leftarrow t^\prime)\Gamma(t^\prime).
\label{divisibility}
\end{equation}
Not all $2\times 2$ stochastic dynamics are divisible, although built out of $2\times2$ divisible matrices. Note that $\Gamma(t^\prime)$ is prescribed by the evolution so, contrary to the case of pure matrices \eqref{matrix_divisibility} where one needs to specify a pair $\hat\Gamma$ and $\tilde\Gamma$, for stochastic dynamics, only the existence of a single stochastic matrix needs to be verified in this second case, namely the transition matrix $\Gamma(t \leftarrow t^\prime)$ that maps probabilities from $t^\prime$ to $t$. This role is clear when composing \eqref{divisibility} with \eqref{Gamma_p_compact}.
\begin{equation}
p(t) = \Gamma(t \leftarrow t^\prime)p(t^\prime)
\label{divisibility_p}
\end{equation}
From this point of view, the Bayesian marginalisation \eqref{Gamma_p_compact} is the requirement of a division event at $t=0$\footnote{The existence of at least one division event is a fundamental assumption required by this formalism. This assumption is not in itself too restrictive and it can be easily justified on physical grounds. For a tabletop experiment, a division event may correspond to a preparation or a measurement of N fully distinguishable quantities. If the system is large or not subject to a preparation, any instant in which it is legitimate to neglect the coherences can be taken as a reference time. The description of the patch corresponding to the entire visible universe with this formalism is not impossible a priori. It would require a foliation and singling out a preferred leaf but this is beyond the scope of this paper.}.
When \eqref{divisibility} holds, the instant $t^\prime$ is called a division event. A stochastic process evolving according to a curve $\Gamma(t)$ in the space of conditional probabilities is called {\it divisible} at time $t^\prime$ if \eqref{divisibility} holds, otherwise it is called {\it indivisible}.

It is by assuming that one can always write \eqref{divisibility_p} for every $t$ and $t^\prime$ that one does not see interferences on classical systems. By dropping it, i.e., not assuming that \eqref{divisibility} holds, it is possible to model interference and entanglement using solely probabilities as these are aspects of an indivisible stochastic evolution.

From a different point of view, divisibility implies a restriction on the possible curves in probability space, meaning that {\it indivisible dynamics is more general than divisible dynamics} and, consequently, able to model a larger class of physical systems which, it turns out, include Quantum systems, open or closed.

However, little is known about divisible evolutions for arbitrary $N$. Buscemi and Datta \cite{Buscemi:2016} provide an informational-theoretic criterion for divisibility in the case of discrete dynamical maps and discrete-time Markov chains (DTMC). Curiously, it is their result about DTMC which, under the stochastic-quantum correspondence, provides the connection with generalised stochastic systems.
Divisibility at $t^\prime$ implies that the stochastic matrix $\Gamma(t^\prime)$ from $0$ to $t^\prime$ is information-decreasing and vice-versa.

A dynamical law prescribes how a system evolves in time, thus specifying how probabilities evolve from an initial time to $t$ via $\Gamma(t)$. This is an important point worth emphasizing: \emph{generalised stochastic systems are associated with a prescribed $\Gamma(t)$ for all valid times}.
From this point of view, when studying the divisibility condition \eqref{divisibility}, it implies that both $\Gamma(t)$ and $\Gamma(t^\prime)$ are given and the question of divisibility of time-evolution is only whether or not a {\it stochastic} matrix $\Gamma(t \leftarrow t^\prime)$ exists such that \eqref{divisibility} is satisfied.

\section{System with two configurations}
\label{sec:two_configurations}

The configurations will be labeled $A$ and $B$ and, fixing $t$ and $t^\prime$, for convenience we shall use the following parametrisation, without any loss of generality.
\begin{align}
  \Gamma(t)
  &\equiv
  \Gamma(t\leftarrow0)
  \\
  &=\begin{pmatrix}
  \Gamma_{AA}\left(t\right) & \Gamma_{AB}\left(t\right)\\
  \Gamma_{BA}\left(t\right) & \Gamma_{BB}\left(t\right)
  \end{pmatrix}\nonumber\\
  &=\begin{pmatrix}
  p\left(A,t| A,0\right) & p\left(A, t| B,0\right)\\
  p\left(B, t A,0\right) & p\left(B, t| B,0\right)
  \end{pmatrix}\nonumber\\
  &=
  \begin{pmatrix}
  p(t) & 1-q(t)\\
  1-p(t) & q(t)
  \end{pmatrix}, \quad 0\leqslant p,q \leqslant1
 \label{Gamma2}
\end{align}

\begin{figure}[!htp]
\centering
\includegraphics{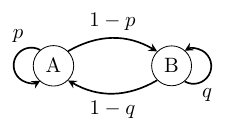}
\caption{The weighted oriented graph associated to the most general transition matrix for a generalised stochastic system with 2 configurations, \eqref{Gamma2}. The two configurations, A and B, have probabilities to remain unchanged after a time lapse $t$ given, respectively, by $p(t)$ and $q(t)$, with $0\leqslant p,q \leqslant1$. It is important to emphasize that the functions $p(t)$ and $q(t)$ are not transition rates but conditional probabilities and that indivisibility prevents conditioning at arbitrary times.}
\label{fig:01}
\end{figure}

At any given moment of time, $\Gamma(t)$ has an associated weighted and oriented graph encoding  the transition probabilities, as shown in figure \ref{fig:01} using the parametrisation \eqref{Gamma2}, with nodes representing configurations and weighted oriented edges representing transition with the respective weights given by the transition probabilities.

\begin{figure}[!htp]
\includegraphics[width=0.35\textwidth]{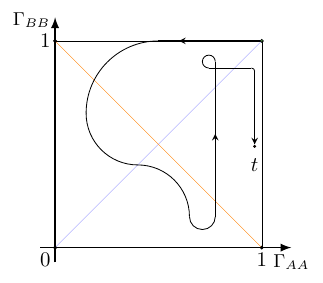}
\caption{An example of an indivisible stochastic evolution for a system with two configurations seen as a curve in the space of stochastic matrices. The stochastic matrix $\Gamma$ is fully specified by its diagonal values, $\Gamma_{AA}^{}$ and $\Gamma_{BB}^{}$ which provide the present choice of coordinates. The evolution from a division event at $t=0$, when the probabilities evolve continuously, is a continuous curve starting at the identity matrix, at $(1,1)$, and moving through the square, the space of $2\times2$ stochastic matrices.}
\label{fig:02}
\end{figure}

Figure \ref{fig:01} displays a fixed-time representation. To depict the time evolution, one may represent all values of $\Gamma(t)$ for all allowed times, by plotting the curved traced by the diagonals inside the $[0,1]^2$ square, i.e., the pair $(\Gamma_{AA}(t), \Gamma_{BB}(t))$, as in figure \ref{fig:02}.

A curve $\Gamma(t)$ in the space of stochastic $2\times2$ matrix is continuous when all its entries are continuous. Also, $\Gamma(t)$ is always left-stochastic if and only iff $(\Gamma_{AA}(t), \Gamma_{BB}(t))$ is a point of $[0,1]^2$ for every possible value of $t$. Therefore, the evolution is continuous and left-stochastic if and only if its curve from $[a, b]$ to $[0,1]^2$  given by $(\Gamma_{AA}(t), \Gamma_{BB}(t))$ is continuous and does not leave the $[0,1]^2$ square of the representation of figure \ref{fig:02}.

{\it Any} continuous curve crossing the secondary diagonal in figure \ref{fig:02} (orange line) corresponds to an indivisible stochastic dynamics. This is a sufficient condition. A necessary and sufficient condition for a divisible evolution is depicted in figures \ref{fig:09} and \ref{fig:10}, valid for any stochastic evolution, continuous or not. For an application of the criteria of figures \ref{fig:09_10} to a continuous curve, see figure \ref{fig:sequence}.

\subsubsection{Degenerate $2\times2$ stochastic matrices}

\begin{figure}[!htp]
\begin{minipage}[h]{0.3\textwidth}
\subfloat[Generic irreversible case.]{
\includegraphics{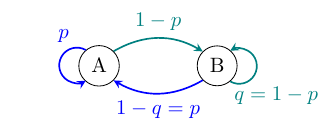}
\label{fig:03}
}\hfill
\end{minipage}
\begin{minipage}[h]{0.3\textwidth}
\subfloat[A is absorbent.]{
\includegraphics{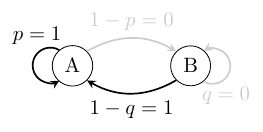}
\label{fig:04}
}
\end{minipage}
\begin{minipage}[h]{0.3\textwidth}
\subfloat[B is absorbent.]{
\includegraphics{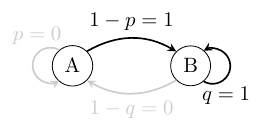}
\label{fig:05}
}
\end{minipage}
\caption{Graphs of degenerate matrices, ($p+q=1$): {\it divisible} dynamics drive the system towards these cases and {\it indivisible} away from them. The conditional probabilities for arriving at a given node are the same, no matter the path. The which-path information is erased and the evolution is irreversible. Graphs \ref{fig:04} and \ref{fig:05} represent the deterministic limits of graph \ref{fig:03} and are related by conjugation by the Pauli matrix $\sigma_x$ which is represented in figure \ref{fig:08}. In section \ref{ssec:dilation_coarse} we show that these are the result of dilations by coarse-graining.}
\label{fig:03_04_05}
\end{figure}

The irreversible matrices correspond to the degenerate case where $p+q=1$ and are all given by the following transition matrix.
\begin{equation}
\label{Gamma2_degenerate}
\Gamma_{degenerate}(t) =
\begin{pmatrix}
p & p\\
1-p & 1-p
\end{pmatrix}, \quad 0\leqslant p \leqslant1
\end{equation}
with the corresponding graph given by figure \ref{fig:03}. This case is characterised by the equal arrival probabilities at any given node, perfectly scrambling the information about the last previous configuration and rendering the evolution irreversible. In other terms, even when one starts in either configuration $A$ or $B$ at $t=0$ with certainty, at time $t$, this information is lost {\it if a division event occurs at $t$.} However, if $t$ is part of a period of indivisible evolution, this may not be the case for posterior times.

The graphs \ref{fig:04} and \ref{fig:05} are the two deterministic limits of graph \ref{fig:03}. Despite being clearly asymmetric, by exchanging the labels $A\longleftrightarrow B$, the graphs \ref{fig:04} and \ref{fig:05} are mapped to each other and their the conditional probability matrices \eqref{Gamma2} reduce to the following {\bf oblique projectors} onto the respective nodes.

\begin{equation}
\label{Gamma2_absorbent}
\Gamma_{\smallselfleft A\leftarrow B}(t) =
\begin{pmatrix}
1 & 1\\
0 & 0
\end{pmatrix}\equiv \Pi_A
,~~
\Gamma_{A\rightarrow B\smallselfright}(t) =
\begin{pmatrix}
0 & 0\\
1 & 1
\end{pmatrix}\equiv \Pi_B
\end{equation}

{\it The general degenerate transition matrix \eqref{Gamma2_degenerate} is a convex combination of the absorbing cases \eqref{Gamma2_absorbent} which are the only deterministic irreversible possibilities.}
\begin{equation}
\label{Gamma2_degenerate_decomposition}
\Gamma_{degenerate} = p \Pi_A + (1-p)\Pi_B, \quad 0\leqslant p\leqslant 1
\end{equation}

We can therefore see the graphs \ref{fig:04} and \ref{fig:05} as a basis of graph \ref{fig:03} through convex combinations. The trajectories obtained from \eqref{Gamma2_degenerate_decomposition} are the same as those one would obtain by, at every passage from $0$ to $t$, one sorts one evolution operator amongst $\Pi_A$ and $\Pi_B$ with respective probabilities $p$ and $1-p$. Each of the projectors \eqref{Gamma2_absorbent} provides a matrix representation of a process exemplified in \cite{Plenio:2001} and discussed in \cite{Kastner:2023} where a molecule which could initially be located at the left (A) or right (B) half of a reservoir, is sent by a piston to the one specific half, with the final position not depending on the initial one\footnote{The representation by a stochastic matrix is the same regardless of the ontological or epistemic character of the probabilities attributed to the molecule's initial position.}.
A degenerate matrix \eqref{Gamma2_degenerate_decomposition} represents, in this scenario, a situation in which there is a probability $p$ that a piston pushes the molecule to $A$ and $(1-p)$ that a piston pushes it to $B$, with all the initial information erased.

The stochastic matrices \ref{Gamma2_degenerate_decomposition} are also projectors. In subsection \ref{ssec:dilation_coarse} we will extend the definition of dilations by coarse graining of \cite{Schmidt:2021} and in subsection \ref{ssec:dilation_degeneracy} we show that the above degenerate cases can be seen as dilations by coarse graining of the trivial, single configuration, case.

\subsubsection{Non-degenerate $2\times2$ stochastic matrices}

\begin{figure}[!htp]
\centering
\begin{minipage}{0.3\textwidth}
\centering
\subfloat[Generic reversible case.]{
\includegraphics{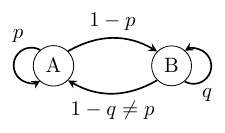}
\label{fig:06}
}
\end{minipage}
\hfill
\begin{minipage}{0.3\textwidth}
\centering
\subfloat[Disjoint evolution.]{
\includegraphics{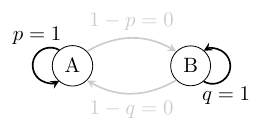}
\label{fig:07}
}
\end{minipage}
\hfill
\begin{minipage}{0.3\textwidth}
\centering
\subfloat[Simple permutation.]{
\includegraphics{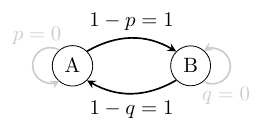}
\label{fig:08}
}
\end{minipage}
\caption{Reversible cases, $p+q\neq1$ : {\it divisible} dynamics drive the system away from these cases and {\it indivisible} towards them. The general reversible scenario (a) has two deterministic limits, (b) a simple permutation, keeping the graph connected and (c) the disjunction of the graph into two trivial nodes. In all these cases, the conditional probabilities for arriving at a given node are different for different paths. A continuous path connecting the identity \ref{fig:07} and the permutation \ref{fig:08} is necessarily {\it indivisible}.}
\label{fig:06_07_08}
\end{figure}

A non-zero determinant for \eqref{Gamma2} defines the non-degenerate case. Figure \ref{fig:08} represents by graph \ref{fig:06} the general two configurations reversible transition matrix and its deterministic limits: the disjoint evolution \ref{fig:07} and the permutation \ref{fig:08}.

The dynamics of a simple permutation represented by graph \ref{fig:08} will be sometimes denoted as $A\rightleftarrows B$. This is the only possible $2\times2$ matrix to represent a reversible deterministic evolution of two connected configurations with discrete time-steps. The conditional probabilities \eqref{Gamma2} reduce to a permutation matrix $\sigma_x$ which is its own inverse.
\begin{equation}
\Gamma_{\smallselfleft\text{A}|\text{B}\smallselfright}(t) = \mathds{1} =
\begin{pmatrix}
1 & 0\\
0 & 1
\end{pmatrix},
\quad
\Gamma_{\text{A}\rightleftarrows \text{B}}(t) = \sigma_x=
\begin{pmatrix}
0 & 1\\
1 & 0
\end{pmatrix}
\label{n2_reversible_deterministic}
\end{equation}
In summary, the only deterministic reversible cases are given by the only two real Pauli matrices with non-negative entries.

While the degenerate $2\times2$ matrices are a convex combination of irreversible deterministic matrices, the most general non-degenerate transition matrix \eqref{Gamma2} {\it is not} a convex combination of the reversible deterministic cases \eqref{n2_reversible_deterministic}. The following decomposition shows that its symmetric part, which is bistochastic, is a convex combination of permutation matrices.
\begin{equation}
\Gamma(t)
=
\sigma_x
+ \left(\frac{p+q}{2}\right)\left(\mathds{1}-\sigma_x\right)
+ \left(\frac{p-q}{2}\right)\left(\Pi_A-\Pi_B\right)
 \label{Gamma2_deterministic_decomposition}
\end{equation}

Increasing $p+q$ decreases the probability of permutation and $\left(\mathds{1}-\sigma_x\right)$ corresponds to precisely this direction in the $2\times2$ matrix space. Increasing $p-q$ directs the system towards the dynamics represented by the graph \ref{fig:04} and away from the dynamics represented by the graph \ref{fig:05} which are both {\it asymmetric graphs}.

\subsection{Divisible and indivisible stochastic dynamics}
\label{ssec:divisible_two}

As emphasized in section \ref{ssec:divisible_and_indivisible}, the determination of
whether or not a system evolves indivisibly is reduced to the existence of a stochastic
matrix $\Gamma(t \leftarrow t^\prime)$, given a pair of stochastich matrices $\Gamma(t)$ and $\Gamma(t^\prime)$, such that \eqref{divisibility}.

If $\Gamma(t)$ is {\it invertible}, then divisibility at time $t^\prime$ implies that {\it both $\Gamma(t^\prime)$ and $\Gamma(t\leftarrow t^\prime)$ are invertible}, since
\begin{equation}
\det\left(\Gamma(t)\right)
=
\det\left(\Gamma(t\leftarrow t^\prime)\right)
\det\left(\Gamma(t^\prime)\right).
\label{divisibility_det}
\end{equation}
Similarly, if $\Gamma(t)$ is {\it non-invertible}, it can only be divisible at time $t^\prime$ if it is the product of two matrices $\Gamma(t^\prime)$ and $\Gamma(t\leftarrow t^\prime)$ with at leat one of these two being non-invertible. The set of equations for divisibility not only naturally splits according to the determinant of $\Gamma(t^\prime)$ but, in fact, the space of non-invertible $2\times2$ stochastic matrices divides the transition probabilities space into two regions and some results depend also on the sign of $\det\Gamma(t^\prime)$.

\subsubsection{$\Gamma(t^\prime)$ is not invertible}

This means that $\Gamma(t^\prime)$ lies on the secondary diagonal of the square \ref{fig:02}, i.e. the one-dimensional subspace of the space of $2\times2$ stochastic matrices with zero determinant which admits the following parametrisation.
\begin{equation}
\Gamma(t^\prime) =
\begin{pmatrix}
r & r\\
1-r & 1-r
\end{pmatrix}.
\end{equation}
As argued above, divisibility implies a relation between determinants that shows that $\Gamma(t)$ needs to be non-invertible, meaning that equation \eqref{divisibility} becomes
\begin{equation}
  \begin{pmatrix}
  p & p\\
  1-p & 1-p
  \end{pmatrix}
=
\begin{pmatrix}
u & 1-v \\
1-u & v
\end{pmatrix}
\begin{pmatrix}
r & r\\
1-r & 1-r
\end{pmatrix}.
\label{non_invertible}
\end{equation}
With the always available choice $(u, v)=(p,1-p)$ which translates to $\Gamma(t\leftarrow t^\prime)=\Gamma(t)$, divisibility is always possible, for every $r$, with no constraints on $\Gamma(t^\prime)$ besides the degeneracy that we already imposed. Other solutions exist for \eqref{non_invertible} but their identification is not necessary for the present task and they are out of the scope of this paper. Therefore, a non-invertible $\Gamma(t)$ may have {\it any other degenerate stochastic matrix} $\Gamma(t^\prime)$ as part of its past and still be divisible. In the next subsection we consider the case when $\Gamma(t^\prime)$ is invertible.

\subsubsection{$\Gamma(t^\prime)$ is invertible}

Stochasticity of $\Gamma(t)$ and $\Gamma(t^\prime)$ together with the invertibility of $\Gamma(t^\prime)$ guarantee that the entries of $\Gamma(t\leftarrow t^\prime)$ sum to one\footnote{This is true for any number of configurations. Let $M_1$ and $M_2$ be two $N\times N$ matrices. Requiring that the columns of $M_1$ and $M_2$ add to one is the same as stating that a row vector of ones is always a left-eigenvector with eigenvalue 1, for $\left(1, \dots, 1\right)M_1=\left(1, \dots, 1\right)M_2=\left(1, \dots, 1\right)$.
As a consequence, whenever $M_\ell$, $\ell=1,2$, is invertible, the columns of $M_\ell^{-1}$ sum to one, since $\left(1, \dots, 1\right)\mathds{1} = \left(1,\dots, 1\right)M_\ell M_\ell^{-1}(t^\prime) = \left(1, \dots, 1\right)M_\ell^{-1}$.
Finally, stochasticity of $\Gamma(t)$ and $\Gamma(t^\prime)$ requires their columns to have unit sum, implying that the columns of $\Gamma(t)\Gamma^{-1}(t^\prime)$ have the same property:
$\left(1,\dots, 1\right)\Gamma(t)\Gamma^{-1}(t^\prime) = \left(1, \dots, 1\right)\Gamma^{-1}(t^\prime)=\left(1, \dots, 1\right)$.
}.
\begin{equation}
\Gamma(t\leftarrow t^\prime) = \Gamma(t)\Gamma^{-1}(t^\prime)
\label{divisibility_inv_long}
\end{equation}
and once it is , $\Gamma(t\leftarrow t^\prime)$ is guaranteed to be stochastic. where left-multiplication by a row of ones amounts to sum the entries of each column.

$\Gamma(t^\prime)$ is parametrized in a similar way to $\Gamma(t)$ in equation \eqref{Gamma2}, but omitting the time-dependence. We shall identify $\Gamma(t^\prime)$ with the point $(r,s)$ in the square representation, i.e.,
\begin{equation}
\Gamma(t^\prime) =
\begin{pmatrix}
r & 1-s\\
1-r & s
\end{pmatrix}.
\end{equation}
with $\det\Gamma(t^\prime)=r+s-1$. Equation \eqref{divisibility_inv_long} becomes
\begin{align}
\Gamma(t \leftarrow t^\prime)
&=
\frac{1}{r+s-1}
\begin{pmatrix}
{ps-(1-q)(1-r)}& {r(1-q)-p(1-s)}\\
{s(1-p)-q(1-r)} & {qr-(1-p)(1-s)}
\end{pmatrix}
\label{Gamma2_t_t_prime}
\end{align}
This matrix is stochastic iff its coefficients are between 0 and 1, which leads, in principle, to 8 inequalities. However, we know that all columns in \eqref{Gamma2_t_t_prime} sum to one, reducing the inequalities to four. Since these inequalities depend on the sign of $\det\Gamma(t^\prime)=r+s-1$, they naturally split according to whether $r+s-1>0$ or $r+s-1<0$.

Under the assumption of $r+s-1>0$ the following expressions valid at the interior of the fundamental square of $\Gamma(t)$, i.e., when $(p, q)\in(0,1)\times(0,1)$.
\begin{subequations}
\begin{align}
&s>1-r
\label{n2_inequalities_4a}
\\
&s\geqslant \max\left(1, \frac{q}{1-p}\right)-\frac{q}{1-p}r
\label{n2_inequalities_4b}
\\
&s \geqslant \max\left(1, \frac{1-q}{p}\right)-\frac{1-q}{p}r
\label{n2_inequalities_4c}
\end{align}
\label{n2_inequalities_4}
\end{subequations}
The inequalities \eqref{n2_inequalities_4} provide lower bounds on $s$ seen as a function of $p$, $q$ and $r$. The case $r+s-1<0$ simply amounts to flipping the inequalities \eqref{n2_inequalities_4} and lead to the upper bounds on $s$ summarised below.
\begin{subequations}
\begin{align}
&s<1-r
\label{n2_inequalities_6a}
\\
&s\leqslant \min\left(1, \frac{q}{1-p}\right)-\frac{q}{1-p}r
\label{n2_inequalities_6b}
\\
&s \leqslant \min\left(1, \frac{1-q}{p}\right)-\frac{1-q}{p}r
\label{n2_inequalities_6c}
\end{align}
\label{n2_inequalities_6}
\end{subequations}

\begin{figure}[!htp]
\begin{minipage}{0.45\textwidth}
\subfloat[p+q-1>0]{
\includegraphics[width=\textwidth]{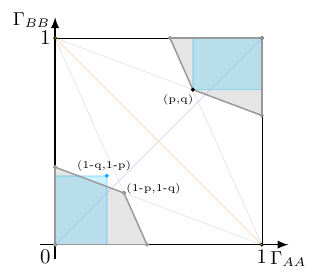}
\label{fig:09}}
\end{minipage}
\hfill
\begin{minipage}{0.45\textwidth}
\subfloat[p+q-1<0]{
\includegraphics[width=\textwidth]{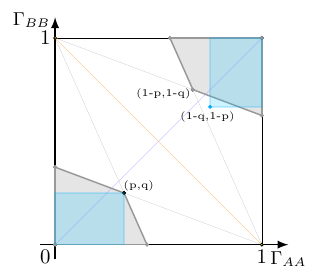}
\label{fig:10}}
\end{minipage}
\caption{Locus of the possible points in the past past of (p,q) which are compatible with a divisibility at time t (gray irregular quadrilaterals) and of the locus of the possible transition matrices (rectangular cyan areas).}
\label{fig:09_10}
\end{figure}

Inequalities \eqref{n2_inequalities_4} and \eqref{n2_inequalities_6} correspond to $(r,s)$ only in the gray regions of figure \eqref{fig:09_10}.

Identitying the regions where $\Gamma(t \leftarrow t^\prime)$ may be found is analogous. When $\Gamma(t \leftarrow t^\prime)$ is invertible, multiply both sides of \eqref{divisibility} by the inverse of $\Gamma(t \leftarrow t^\prime)$ and impose the positivity of the right-hand side. In practice, let $\Gamma(t \leftarrow t^\prime)$ be parametrised by its diagonal entries $(u, v)$. We need to ensure the positivity of
\begin{equation}
\Gamma(t^\prime)
=
\frac{1}{u+v-1}
\begin{pmatrix}
p+u-1 & v-q\\
u-p & q+u-1
\end{pmatrix}
\end{equation}
which amounts to locate $\Gamma(t \leftarrow t^\prime)$ in the blue regions of figure \ref{fig:09_10}. The treatment of the case when $\Gamma(t \leftarrow t^\prime)$ is not invertible is a trivial consequence of the above discussion for the non-invertible cases and it is solved for $u=p, v=1-p$, thus establishing that when $(p,q)$ becomes degenerate, the blue regions combine and intersect only at the point $(p, 1-p)$. The two disjoint gray areas degenerate to two disjoint line segments or points when $(p,q)$ is at an edge or a vertex of the main diagonal, respectively.

When $p$ approaches $1-q$, i.e., when $\det\Gamma(t)$ approaches zero, conditions \eqref{n2_inequalities_4b} and \eqref{n2_inequalities_4c} approach $s\leqslant 1-r$, seeming to contradict condition \eqref{n2_inequalities_4a}. Similarly for \eqref{n2_inequalities_4b} and \eqref{n2_inequalities_4c} which include the lower border for f while \eqref{n2_inequalities_4a} excludes it. The fact that we proved that all degenerate $\Gamma(t)$, those that lie on the secondary diagonal, are divisible with $\Gamma(t^\prime)$ anywhere in the square shows that when $p+q=1$, the two gray regions shown in figure \ref{fig:09} for $p+q>1$ and the two corresponding ones in \ref{fig:09} for $p+q<1$ get fused to fill the entire square.

The form of the inequalities \eqref{n2_inequalities_4} and \eqref{n2_inequalities_6} suggest a special treatment when $p$ goes to zero or one. In these cases, a simple calculation shows that divisibility fixes the evolution to remain at the same boundary and that the past of $\Gamma(t)$ is between $\Gamma(t)$ and $\mathds{1}$ for $p=1$ or $\sigma_x$ for $p=0$, exactly what should be expected from the limit of the gray areas as $p$ tends to $1$ or $0$. Therefore, this geometric construction provides the full picture of stochastic divisible evolutions for two configurations.

To every point with coordinates $(r,s)$ in the gray regions, equation \eqref{Gamma2_t_t_prime} associates a point $(u,v)$ given by the diagonal of $\Gamma(t \leftarrow t^\prime)$. Under this relation, the gray areas generate the cyan areas which show where $\Gamma(t \leftarrow t^\prime)$ may be found given a $\Gamma(t)$ with diagonal $(p,q)$.

The symmetry arguments of subsection \ref{ssec:symmetries} are clearly visible in figures \ref{fig:09_10}. From equation \eqref{divisibility_symmetry}, to every ordered pair $(\Gamma(t\leftarrow t^\prime), \Gamma(t^\prime))$ that divide $\Gamma(t)$ we may associate other ordered pairs given by $(\Gamma(t\leftarrow t^\prime)\sigma^{-1}, \sigma\Gamma(t^\prime))$ where $\sigma$ is a permutation matrix, here there are only two permutations matrices, the Pauli matrix $\sigma_x$ and the identity, and both are their own inverses. The right-action of $\sigma_x$ permutes the columns and its left action permutes the rows, explaining the matching of the blue and gray regions at $(p,q)$ and their mismatch at the other side of the square.

In subsection \ref{ssec:density_connectivity}, by simulating regions \ref{fig:09_10} we obtain information regarding the relative position of the pairs $(\Gamma(t\leftarrow t^\prime), \Gamma(t^\prime))$ and the density of $\Gamma(t\leftarrow t^\prime)$ when $\Gamma(t^\prime)$ is uniformly drawn. We also analyse the symmetry aspects of some features in the light of subsection \ref{ssec:symmetries}.

Having identified where the divisors of $\Gamma(t)$ may be located, it is time to consider the {\it future} of $\Gamma(t)$ under divisible and indivisible evolution, i.e., the set of stochastic matrices that is reachable from $\Gamma(t)$ under different types of dynamics. We shall see that divisibility of the {\it process} puts the strongest constraint and brings the greatest degree of similarity with a relativistic setting.

\begin{figure}[!htp]
\includegraphics[width=0.45\textwidth]{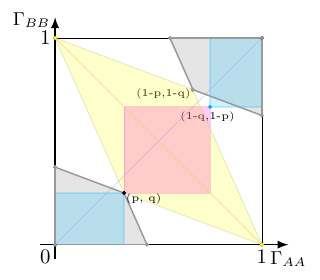}
\includegraphics[width=0.45\textwidth]{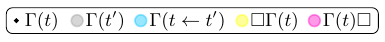}
\caption{Summary of all the regions presented in this section for a stochastic matrix with diagonal entries (p,q).}
\label{fig:11_12}
\end{figure}

Suppose that a division event happens at time $t$. This means, as defined in \cite{Barandes:2023ivl} that for $\tilde t \geqslant t$ there is a matrix $\Gamma(\tilde t \leftarrow t)$ such that
\begin{equation}
\Gamma(\tilde t)=\Gamma(\tilde t \leftarrow t)\Gamma(t).
\label{divisibility_forwards}
\end{equation}
If all we know is $\Gamma(t)$, we can determine what are the possible $\Gamma(\tilde t)$ if we allow the transition matrix $\Gamma(\tilde t \leftarrow t)$ to be {\it any} $2\times2$ stochastic matrix. In other words, to obtain the possible values of $\Gamma(\tilde t)$ one should left-multiply $\Gamma(t)$ by every point in the square. This set of matrices will called {\it the square on $\Gamma(t)$} and denoted by $\square\Gamma(t)$.
\begin{align}
\square\Gamma(t)
=&~ \Bigg\{
\begin{pmatrix}
xp +(1-y)(1-p) & x(1-q)+q(1-y)\\
(1-x)p+(1-p)y & yq + (1-x)(1-q)
\end{pmatrix}
\nonumber\\
&
\qquad\qquad\Big|~
(x,y)\in [0,1]^2
\Bigg\}
\nonumber\\
=&~\left\{H\Gamma(t)~|~ H\in \square\right\}
\end{align}
When $\Gamma(t)$ is invertible, each element of $\square\Gamma(t)$ is a $2\times2$ stochastic matrix $W$ such that $W\left(Gamma(t)\right)^{-1}$ is stochastic. This is precisely the condition which makes this problem similar to the one of the transition matrix \eqref{Gamma2_t_t_prime}. If $(w, z)$ are the diagonal entries of $W$, this amounts to
\begin{align}
\frac{1}{p+q-1}
\begin{pmatrix}
{qw-(1-p)(1-z)}& {p(1-z)-w(1-q)}\\
{q(1-w)-z(1-p)} & {pz-(1-w)(1-q)}
\end{pmatrix}\geqslant0
\nonumber
\end{align}
where positivity should hold entry-wise. Again, this splits into two cases depending on the sign of $\det\Gamma(t)=p+q-1$ and leads, in each case, to four inequalities. After separately considering the boundary cases, the result is an explicit region for the points $(w,z)$ in the square which belong to $\square\Gamma(t)$.
\begin{subequations}
\begin{align}
&w\leqslant \max\left(1, \frac{q}{1-p}\right)-\frac{q}{1-p}z
\label{future_inequality_a}
\\
&w \leqslant \max\left(1, \frac{1-q}{p}\right)-\frac{1-q}{p}z
\label{future_inequality_b}
\\
&w\geqslant \min\left(1, \frac{q}{1-p}\right)-\frac{q}{1-p}z
\label{future_inequality_c}
\\
&w \geqslant \min\left(1, \frac{1-q}{p}\right)-\frac{1-q}{p}z
\label{future_inequality_d}
\end{align}
\label{future_inequalities}
\end{subequations}
These inequalities are simple the reverse of inequalities \eqref{n2_inequalities_4} and \eqref{n2_inequalities_4}. The region $\square\Gamma(t)$ is depicted as the yellow area in figure \ref{fig:11_12}, with the grey area representing, as before, the possible points $\Gamma(t^\prime)$.

Figure \ref{fig:11_12} summarizes all the area studied in this section: he evolution is divisible at $t^\prime<t$ for $\Gamma(t^\prime)$ in the areas bounded by the gray regions; the possible futures of $\Gamma(t)$ correspond to  $\square\Gamma(t)$ given a division event at $t$ and are bounded by $\Gamma(t)$, $\sigma\Gamma(t)$, $\Pi_A$ and $\Pi_B$ as defined in equation \eqref{Gamma2_absorbent}; possible locations for $\Gamma(t\leftarrow t^\prime)$ are given by the blue rectangles near the identity and $\sigma_x$; the central magenta square represents the mapping of the square under $\Gamma(t)$. The situation when $\det\Gamma(t)\leqslant0$ is obtained through a rotation by $\pi$, as in figures \ref{fig:09_10}.

As before, when $p=0$ or $p=1$ two of the lines become vertical, which follows more explicitly from the geometric construction of rays emanating from the deterministic and irreversible matrices and intercepting $\Gamma(t)$ to produce the separation between the past near a deterministic and reversible matrix and future towards the degenerate cases.

Knowing the possible points in the past or future of $\Gamma(t)$ assuming a division event at $t$, there is one missing element to complete the geometric separation of regions. There is a mismatch in figures \ref{fig:09_10} between one of the blue regions where $\Gamma(t\leftarrow t^\prime)$ may be found and the region where $\Gamma(t^\prime)$ may be found. In particular, the vertex at $(1-q, 1-p)$ seems to remain isolated despite the cone around $(p,q)$ and the one aroung $(1-p, 1-q)$ enclosing a region in the center corresponding to the possible values of $X\Gamma(t)$ for as stochastic matrix $X$. Figure  \ref{fig:11_12} closes this gap by illustrating the set that bridges this gap: the result of the action of $\Gamma(t)$ on the whole square which we will interpret after defining it.
\begin{align}
\Gamma(t)\square
=&~\left\{\Gamma(t)H~|~ H\in \square\right\}
\nonumber\\
=&~ \Bigg\{
\begin{pmatrix}
x(p+q-1)+1-q & y(1-p-q)+p\\
x(1-p-q)+q & y(p+q-1)+1-p
\end{pmatrix}
\nonumber\\
&
\qquad\qquad\Big|~
(x,y)\in [0,1]^2
\Bigg\}
\end{align}

Figure \ref{fig:11_12} shows the set $\Gamma(t)\square$ together with the other regions already identified. When $\Gamma(t)$ is divisible at $t=t^\prime$, and $H$ is a stochastic matrix,
\begin{align}
\Gamma(t)H
=\Gamma(t\leftarrow t^\prime)\left[\Gamma(t^\prime)H\right]
\end{align}
is also a stochastic matrix. Varying $H$ inside the full square produces $\Gamma(t)\square$ on both sides, {\it regardless of the value of $\Gamma(t\leftarrow t^\prime)$}. The reason $\Gamma(t)\square$ is related to the set of possible $\Gamma(t\leftarrow t^\prime)$ is that this set contains $\Gamma(t)$ itself and $\Gamma(t)\sigma_x$, the blue vertices with the smallest distance to the center within the blue regions. In the first case, $\Gamma(t^\prime)H$ is simply $H$ since the transition matrix is the identity and in the second case $\Gamma(t^\prime)H$ is  $\sigma_xH$. In other terms, the vertices of the cyan rectangles are precisely the points for which $\Gamma(t)\square=\Gamma(t\leftarrow t^\prime)\square$.

\subsection{Continuous and discontinuous dynamics}
\label{ssec:continuous_two}

Each one of figures \ref{fig:09_10} displays two regions of allowed matrices $\Gamma(t^\prime)$ at the past of the given $\Gamma(t)$. The first situation, figure \ref{fig:09}, has $\Gamma(t)$ has strictly positive determinant and the point $(p, q)$ has the identity, point $(1,1)$,  in its past (grey area). If all portions of the curve for $0\leqslant t^\prime \leqslant t$ are in the closed grey region containing $(p, q)$ and the identity, then the evolution is divisible and it is allowed to be continuos. However, if with $\det\Gamma(t)=p+q-1<0$ the points of $\Gamma(t^\prime)$ for $0\leqslant t^\prime \leqslant t$ will only provide a continuous and divisible curve if they belong to the lower gray region which includes the permutation but not the identity. Consistency of the probabilities at time $t^\prime=0$ requires $\Gamma(0)=\mathds{1}$ so for $\det\Gamma(t)<0$ it is impossible to require, for all times, continuity and divisibility. It is possible, however to choose one. An example of indivisible but continuous evolution is quantum 2-level system oscillating between two orthogonal states under unitary evolution to which we associate two configurations. A discontinuous and divisible evolution reaching the lower region is a discrete-time Markov chain with transition matrix $\sigma_x$. For other interesting cases see section \ref{sec:information}. In general, as long as jumps between regions of opposite determinant (separated by the orange lines in the squares) are allowed, one may have divisibility while exploring the gray region at the lower left\footnote{See subsection \ref{ssec:density_connectivity} for examples of the relative position of the divisors in these different situations, subsection \ref{ssec:symmetries} for the symmetry aspects and \ref{sec:information} for an explicit derivation of their relative position using a coordinate system motivated by the above classification using graphs and by information reduction.}.

\begin{figure}[!thp]
\centering
\includegraphics[width=0.45\textwidth]{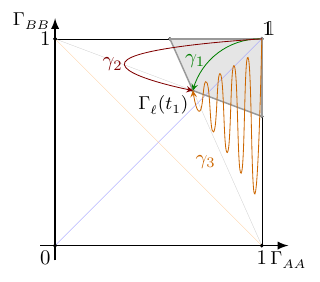}
\caption{Paths $\gamma_\ell$ given by $\Gamma_\ell(t)$ for $\ell=1,2,3$ rom $t=0$ to $t=t_1$ with the value, $\Gamma(t)$, at time $t$. The curves show that knowledge of $\Gamma(t)$ at a chosen time is not sufficient for divisibility of the corresponding curve, since its past must rely in the grey zone for divisibility.}
\label{fig:13}
\end{figure}

From figures \ref{fig:09_10} it is clear that when past of $(p,q)$ is not in the grey region, it does not matter if the curve is continuous or not: the evolution is {\it necessarily indivisible}. The converse is also true. An evolution between times 0 and $t$ is indivisible for if and only if there is at least one value of $\Gamma(t)$ for which its past since $t=0$ is not in the instantaneous past-cone. When the curve is continuous, it may be convenient in some cases to consider the future of a point, represented in yellow in figure \ref{fig:11_12}. Figure \ref{fig:13} illustrates continuous three curves which coincide at time $t_1$ but only $\gamma_1$ is associated with a divisible evolution.

\begin{figure}[!htp]
\centering
\begin{minipage}{0.23\textwidth}
\subfloat[Path traced from $t=0$ to $t=t_1$.]{
\includegraphics[width=\textwidth]{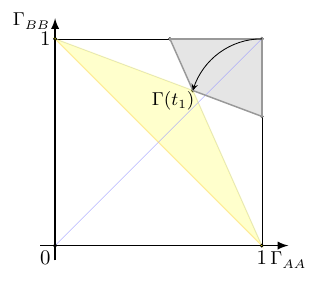}
\label{fig:14}
}
\end{minipage}
\begin{minipage}{0.23\textwidth}
\subfloat[Continuation of the path until $t_2$.]{
\includegraphics[width=\textwidth]{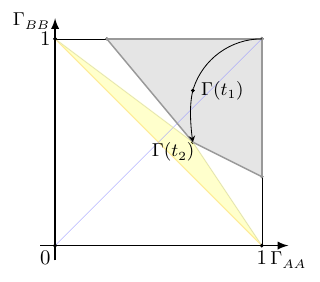}
\label{fig:15}
}
\end{minipage}
\begin{minipage}{0.23\textwidth}
\subfloat[Path until $t_3$, where the determinant is 0.]{
\includegraphics[width=\textwidth]{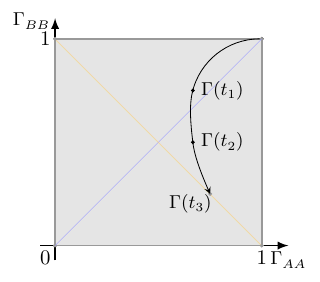}
\label{fig:16}
}
\end{minipage}
\begin{minipage}{0.23\textwidth}
\subfloat[From $t_3$ to $t_4$ the evolution is indivisible.]{
\includegraphics[width=\textwidth]{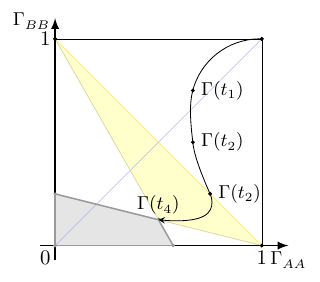}
\label{fig:17}
}
\end{minipage}
\caption{An example of a path $\Gamma(t)$ traced on the space of $2\times 2$ stochastic matrices. The curve starts at the identity, at $(1,1)$ and evolves divisibly during the first three frames, since its past is inside the grey area. The moment it crosses the secondary diagonal at $t_3$ is the last time that it has a full divisible past. At time $t_4$, figure \ref{fig:17}, the grey area is in no longer in its past, so the portion of the evolution from $t_3$ to $t_4$ is indivisible. The yellow areas indicate the possible futures of each point provided the curve is continuous and divisible at every instant.}
\label{fig:sequence}
\end{figure}

Consider the situation from the point of view of the transition matrix $\Gamma(t\leftarrow t^\prime)$ using figure \ref{fig:10} where $(p,q)$ is below the secondary diagonal, i.e., $\det\Gamma(t)=p+q-1<0$. Divisibility at $t^\prime$ implies the existence of $\Gamma(t\leftarrow t^\prime)$. Taking $t^\prime$ to $t$, it approaches $\mathds{1}$ by continuity\footnote{Consistency requires {\it only} that $\Gamma(t\leftarrow t)=\mathds{1}$.} and, consequently, must be in the blue region connected with the identity, but $\Gamma(t)$ is not. If the evolution is continuous for {\it every} $t^\prime\in[0,t]$, then $\Gamma(t\leftarrow t^\prime)$ is also valid as $ t^\prime$ tends to zero, in which case it should equal $\Gamma(t)$. However, divisibility imposes its determinant does not change sign. Therefore, when $\det\Gamma(t)<0$, {\it at most a portion} of the evolution will be continuous {\it and} divisible but never the full path from the identity. This implies, in particular, that any system described by a $\Gamma(t)$ evolving continuously in time will have at least a period of indivisible evolution if it crosses the line $\Gamma_{AA}+\Gamma_{BB}=1$ where its determinant vanishes or if the past of any of the points of $\Gamma(t)$, for any allowed $t$, is outside the closed grey regions.

This is illustrated by the sequence of figures \ref{fig:sequence}. The continuity of the curve $\Gamma(t)$ in matrix space is translated in the continuity of the diagonal terms that provide the coordinates of figures \ref{fig:sequence}.  Compare them with figure \ref{fig:09_10} that shows that, given $\Gamma(t)$, there are two allowed regions where $\Gamma(t^\prime)$ could be, so that the evolution $\Gamma(t)$ is divisible at $t^\prime<t$. Continuity necessarily excludes at least of the regions as a possibility, except when the matrix is degenerate, in which case the entire square can be in its past.

This implies, in particular, the existence of infinitely many stochastic evolutions that are divisible for a certain duration after departing from the identity. Similarly, once $\Gamma(t)$ distances itself from the permutation matrix $\sigma_x$ located at the origin, it will enjoy a period of divisible stochastic evolution under the equivalent circumstances that follow by symmetry.

These distinctions rely solely on continuity and do not require differentiability\footnote{One could, for instance, choose stochastic dynamics by randomly selecting the value of $\Gamma(t)$ from a stochastic process. A valid choice is given by $\Gamma_{11}(t)=\Gamma_{22}(t)=1-\exp(-[x(t)]^2)$, where $t\geqslant0$ and $x(t)$ is a path sampled from a Wiener process. Because such a function $x(t)$ is continuous but almost surely nowhere differentiable, a $\Gamma(t)$ chosen this way would have the same properties.}. It is important to note that the yellow areas represented in figures \ref{fig:sequence} do not form a quadrilateral as in figure \ref{fig:11_12} because it represents a case of a divisible stochastic {\it process} which, as explained in the introduction, must be divisible for all values of $t$ compatible with the dynamics. This is precisely what is required for Markovian systems where providing the probabilities at any given time suffices to determine the entire evolution of the probabilities.

\begin{figure*}[!htp]
\begin{minipage}{0.45\textwidth}
\subfloat[Connectivity of the divisors of $\Gamma(t)$]{
\includegraphics[width=\textwidth]{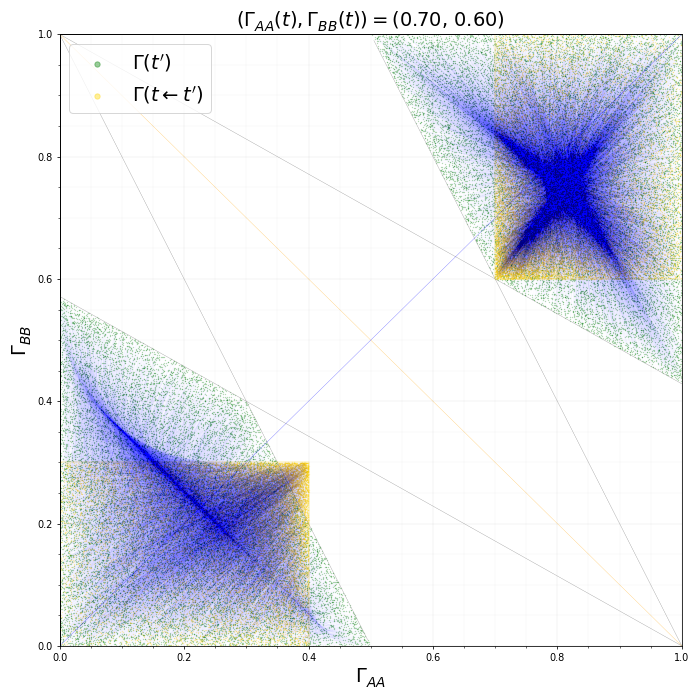}
\label{fig:18}
}
\end{minipage}
\begin{minipage}{0.45\textwidth}
\subfloat[Relative density of the divisors of $\Gamma(t)$.]{
\includegraphics[width=\textwidth]{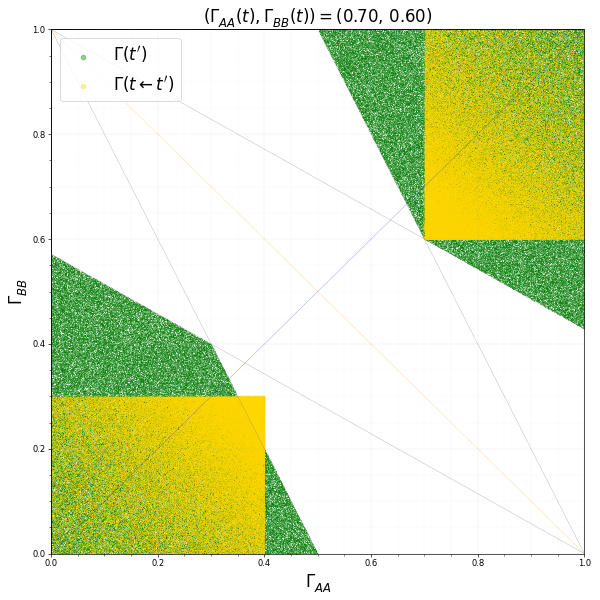}
\label{fig:19}
}
\end{minipage}
\caption{$\Gamma(t)$ with diagonal $(p,q)$ and its divisors. The green points represent possible values of $\Gamma(t^\prime)$ and the yellow points possible values of $\Gamma(t\leftarrow t^\prime)$ such that the ordered pair $(\Gamma(t\leftarrow t^\prime), \Gamma(t^\prime))$ that multiply to $\Gamma(t)$. The blue lines join each ordered pair appearing in figure \ref{fig:18}. The points $\Gamma(t^\prime)$ are uniformly drawn in the allowed region, and $\Gamma(t\leftarrow t^\prime)$, computed, given a $\Gamma(t)$ that they divide according to equation \eqref{divisibility}. The distribution of $\Gamma(t\leftarrow t^\prime)$ accumulates closer to the centre of the square as shown in figure \ref{fig:18}. $\text{diag}(\Gamma(t))=(0.7, 0.6)$}
\label{fig:connectivity_up}
\end{figure*}

\begin{figure*}[!htp]
\begin{minipage}{0.45\textwidth}
\subfloat[Connectivity of the divisors of $\Gamma(t)$]{
\includegraphics[width=\textwidth]{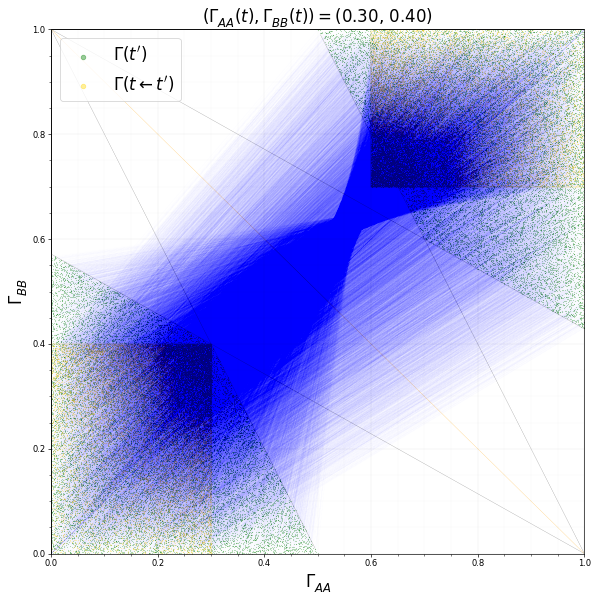}
\label{fig:p3q4_lines}
}
\end{minipage}
\begin{minipage}{0.45\textwidth}
\subfloat[Relative density of the divisors of $\Gamma(t)$.]{
\includegraphics[width=\textwidth]{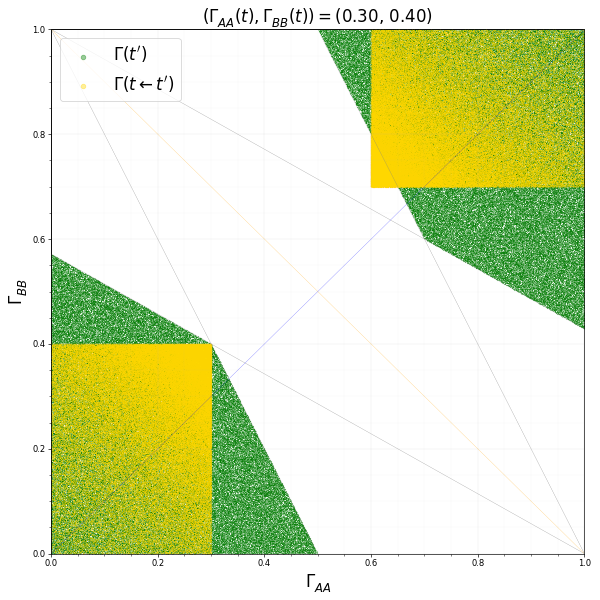}
\label{fig:p3q4}
}
\end{minipage}
\caption{The analogous of figure \ref{fig:connectivity_up} for $\Gamma(t)$ at the opposite side of the square with respect to the center. figure \ref{fig:p3q4_lines} displays the connectivity and \ref{fig:p3q4} the relative densities. The choice $\text{diag}(\Gamma(t))=(0.3, 0.4)$ guarantees that the green areas are the same as those of figure \ref{fig:connectivity_up}, but not the yellow areas.}
\label{fig:connectivity_down}
\end{figure*}

\begin{figure*}[!htp]
\begin{minipage}{0.45\textwidth}
\subfloat[$\text{diag}(\Gamma(t))=(0.2, 0.6)$]{
\includegraphics[width=\textwidth]{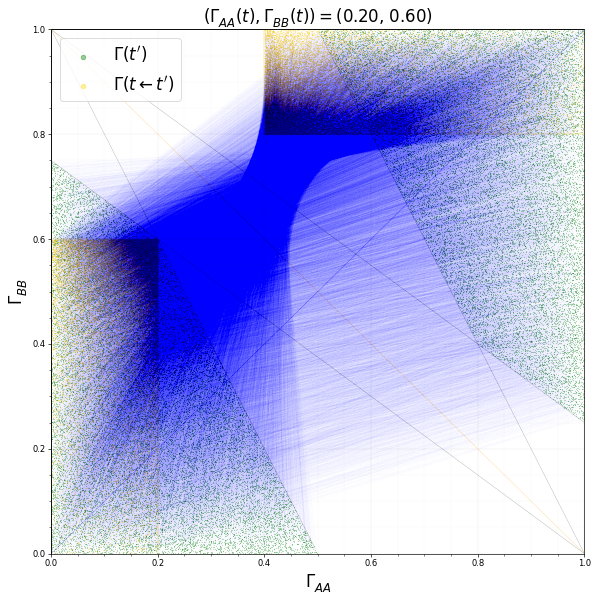}
}
\end{minipage}
\begin{minipage}{0.45\textwidth}
\subfloat[$\text{diag}(\Gamma(t))=(0.8, 0.4)$]{
\includegraphics[width=\textwidth]{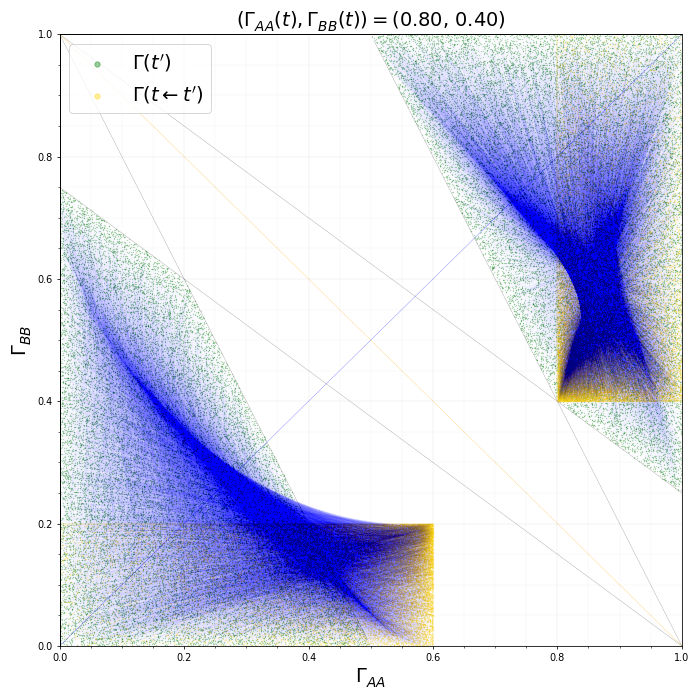}
}
\end{minipage}
\caption{A $\Gamma(t)$ that is closer to the permutation of two elements at the origin than to the identity and its divisors. The green points represent possible values of $\Gamma(t^\prime)$ and the yellow points possible values of $\Gamma(t\leftarrow t^\prime)$ such that the ordered pair $(\Gamma(t\leftarrow t^\prime), \Gamma(t^\prime))$ that multiply to $\Gamma(t)$. The blue lines join each ordered pair appearing in the figure.}
\label{fig:connectivity_2}
\end{figure*}

\subsection{Density and connectivity}
\label{ssec:density_connectivity}

The previous analytic results show the existence of two kinds of regions, one where $\Gamma(t^\prime)$ may be found and another containing each associated $\Gamma(t\leftarrow t^\prime)$. The points in these regions being in a one-to-one correspondence except when $\Gamma(t)$ is degenerate, if we randomly select a $\Gamma(t^\prime)$ in the allowed region,  $\Gamma(t\leftarrow t^\prime)$ will be fixed and in the following figures we illustrate how the pairs are connected and use the symmetry analysis of subsection \ref{ssec:symmetries} to explain them. We also illustrate that $\Gamma(t\leftarrow t^\prime)$ tends to be closed to the center than to the identity, in the sense that a uniformly distribution for  $\Gamma(t^\prime)$ produces a distribution for $\Gamma(t\leftarrow t^\prime)$ that accumulates towards the center of the square, within the bounds of the allowed region.

Figure \ref{fig:connectivity_up} illustrates how pairs of matrices that multiply to $\Gamma(t)$ are connected, figure \ref{fig:18} and distributed, figure \ref{fig:19}.. When $\Gamma(t)$ is closer to the identity than to the permutation, i.e., $\Gamma_{AA}(t)+\Gamma_{BB}(t)>1$, as displayed in figure \ref{fig:connectivity_up}, if  $\Gamma(t^\prime)$ is in the upper region, so is $\Gamma(t\leftarrow t^\prime)$. Also, the pairs are opposed and their lines cross. The green and yellow regions that are not connected to $\Gamma(t)$, lying closer to the permutation matrix at the origin, can be obtained by a right and a left action by permutations\footnote{See subsections \ref{ssec:symmetries} and \ref{ssec:divisible_two}.} which, acting differently on $\Gamma(t^\prime)$ and $\Gamma(t\leftarrow t^\prime)$, undo the crossing.
figure \ref{fig:19} illustrates the difference of densities, with $\Gamma(t^\prime)$ uniformly drawn and $\Gamma(t\leftarrow t^\prime)$ computed to satisfy the divisibility equation \eqref{divisibility}.

Figures \ref{fig:connectivity_2} show two other examples with different values of $\Gamma(t)$ but associated with the same pair of allowed regions for $\Gamma(t^\prime)$. Similar remarks apply regarding the connectivity and the opposing positions of $\Gamma((t\leftarrow t^\prime)$ and $\Gamma(t^\prime)$ when $\Gamma(t^\prime)$ is in the green area connected to $\Gamma(t)$. The flatness of the quadrilaterals is what induces crossings which we would not see in a more symmetric case.

Figure \ref{fig:connectivity_down} shows an example when $\Gamma(t)$ is closer to the permutation, at the origin, than to the identity at $(1,1)$. The most recent divisible part of the past of $\Gamma(t)$ must be in the connected region when $\Gamma(t)$ is continuous. The transition matrix $\Gamma(t\leftarrow t^\prime)$ must be, in this case, in the component connected to the identity as required by continuity, since the limit of $t^\prime$ tending to $t$ sends $\Gamma(t\leftarrow t^\prime)$ to the identity in a continuous evolution. The symmetry arguments of subsection \ref{ssec:symmetries} apply and there is a corresponding relation between opposite areas also for discontinuous dynamics. The overall connectivity is displayed in figure \ref{fig:p3q4_lines}.

Again, each point $\Gamma(t^\prime)$ that is in the allowed region that is connected to $\Gamma(t)$ is associated with a matrix $\Gamma(t\leftarrow t^\prime)$ that is in the opposite side of its associated yellow region, leading to lines between regions that cross each other. For each of these pairs there is a another pair, $\Gamma(t\leftarrow t^\prime)\sigma_x$ and $\sigma_x\Gamma(t^\prime)$ so that transforming the set of crossing lines produces another set for which the lines do not cross.

The relative location of $\Gamma(t\leftarrow t^\prime)$ and $\Gamma(t^\prime)$ given $\Gamma(t)$ can also be understood in terms of equation \eqref{divisibility_det} relating the three determinants.

Regarding the density, the example of section \ref{ssec:ensembles} shows a way to relate divisibility may with an exponential number of repeated operations to reach degeneracy starting from a permutation matrix, if one remains at a connected component. The distributions in the plots above illustrate how this may extend for the cases out of the boundary of the square. The cases of discontinuous divisible evolution with jumps between the green areas follow from the analysis of subsection \ref{ssec:symmetries}, specifically from equation \ref{divisibility_symmetry_T} with a time-dependent choice of intertwining permutation matrix $\sigma$, and section \ref{sec:information} on information decrease.

\subsection{Symmetries, divisibility and continuity}
\label{ssec:symmetries}

In any dimension, divisible and indivisible stochastic matrices never come alone. Divisibility, in particular, implies the existence multiple divisors in the case, as we will see below, when continuity is not required. Also, regardless of (in-)divisibility, continuity of the stochastic {\it dynamics} restrict restricts possible symmetry transformations.

Let $\Gamma$ be a stochastic matrix and $\sigma$ be a permutation matrix\footnote{A permutation matrix is a deterministic stochastic matrix with no empty rows.}. Then $\sigma\Gamma$ is also a stochastic matrix, differing from $\Gamma$ by the position of the entries in each column (permutation of the outputs) . In other words, the left-action of $\sigma$ on $\Gamma$ does change the locations of the columns of $\Gamma$ but only permutes rows (permutation of the inputs) . Also, $\Gamma\sigma$ is a stochastic matrix, differing from $\Gamma$ by an exchange of full columns, keeping each column intact except for its position in the matrix. Therefore, the transformation $T_{(\sigma_1, \sigma_2)} : \mathcal{S}_N\to \mathcal{S}_N$
\begin{equation}
T_{(\sigma_1, \sigma_2)}(\Gamma)=\sigma_2\Gamma \sigma_1^{-1},
\label{permutation_map}
\end{equation}
with permutation matrices $\sigma_1$ and $\sigma_2$, maps stochastic matrices to stochastic matrices, divisible or not. It is easy to see that $T_{(\sigma_3, \sigma_4)}\circ T_{(\sigma_1, \sigma_2)}=T_{(\sigma_3\sigma_1, \sigma_4\sigma_2)}$ and that they form a group. The diagonal subgroup of these transformations, i.e., the subgroup of conjugation transformations $T_{(\sigma, \sigma)}$, for some permutation matrix $\sigma$, is the set of transformations preserving the identity matrix. The diagonal transformations represent a relabelling of the configurations.

A {\it continuous} stochastic {\it dynamics} given by a curve $\Gamma(t)$ is mapped to another continuous stochastic dynamics $\hat\Gamma(t)$ under $T_{(\sigma, \sigma^\prime)}$ if and only if $\sigma=\sigma^\prime$. In other words, a continuous stochastic dynamics is specified up to a relabelling of the configurations. Otherwise, we would either have $\Gamma(0)=\mathds{1}$ and $\hat\Gamma(0)=\sigma_2 \sigma_1^{-1}\neq\mathds{1}$ if $\sigma_2\neq\sigma_1$ or we would need to require that $\hat\Gamma(t)=T_{(\sigma_1, \sigma_2)}(\Gamma(t))$ with $\sigma_2\neq\sigma_1$ for $t\neq0$ and $\sigma_2=\sigma_1$ for $t=0$, introducing a discontinuity since two distinct permutation matrices are disjoint in the space of matrices, with a finite distance\footnote{We use the Euclidean norm of $\mathbb{R}^{N^2}$ for computing distances with each matrix entry associated with a direction in $\mathbb{R}^{N^2}$. Consequently, the scalar product is the Euclidean one, $(A, B)=\text{tr}(A^TB)$.}. Dropping the identity requirement $\hat\Gamma(0)=\mathds{1}$ would create an inconsistency in the initial probability distribution.

Permutations are the only stochastic matrices with inverses that are also stochastic. They are deterministic and do not decrease information\footnote{{\it Irreversible} deterministic stochastic matrices are those with repeated columns, as this repetition implies a zero determinant and a non-trivial kernel. They are, up to permutations, projectors onto proper subsets of configurations.}.

Let  $\Gamma$ be a {\it divisible} stochastic matrix $\Gamma$ and an ordered pair $(\hat\Gamma, \tilde\Gamma)$ of left and a right divisors of $\Gamma$, i.e., the triple $(\Gamma, \hat\Gamma, \tilde\Gamma)$ obeys equation \ref{matrix_divisibility}. Then, for any permutation matrix $\sigma$, the pair $(\hat\Gamma\sigma^{-1}, \sigma\tilde\Gamma)$ also divide $\Gamma$.
\begin{equation}
\Gamma = \hat\Gamma\sigma^{-1}\sigma \tilde\Gamma
\label{matrix_divisibility_symmetry}
\end{equation}
The matrix $\sigma\tilde\Gamma$ differs from $\tilde\Gamma$ by an exchange of rows (permutation of the outputs) and $\hat\Gamma\sigma^{-1}$ from $\hat\Gamma$ by an exchange of columns (permutation of the inputs). As there are $N!$ permutations of $N$ configurations, identifying a pair of divisors of $\Gamma$ implies in the identification of $N!$ pairs of divisors.

Composing \eqref{matrix_divisibility_symmetry} with \eqref{permutation_map} we can obtain up to $(N!)^2$ divisible matrices from a given $\Gamma$, each with $N!$ pairs of divisors. This is an upper limit since there may be cases where many of these matrices coincide. For instance, if $\Gamma$ is the centre of the stochastic polytope, i.e., $\Gamma_{ij}=1/N$, for every $i, j\in\{1,\dots, N\}$, permutations will keep this matrix intact. A similar statement holds for subsets $\mathcal{C}_k$ of the set of configurations $\mathcal{C}$: if $\mathcal{C}_k$ has size $k$ and $\Gamma_{ij}=1/k$ for $i, j\in\mathcal{C}_k$, then permutations acting solely on $\mathcal{C}_k$ will keep $\Gamma_{ij}$ intact.

A similar argument holds for divisible {\it dynamics}, with the identity inserted as $\sigma^{-1}\sigma$ at the right hand side of equation \ref{divisibility}.
\begin{equation}
\Gamma(t) = \Gamma(t \leftarrow t^\prime)\sigma^{-1}\sigma\Gamma(t^\prime)
\label{divisibility_symmetry}
\end{equation}

The only difference comes in the composition of the ordered pairs of divisors of $\Gamma(t)$ with the transformation \eqref{permutation_map} which, as explained above, must be limited to the diagonal transformations or conjugations if continuity holds. At first sight, it is not immediate that the divisors should also be conjugated in \eqref{divisibility_symmetry} since the action of a diagonal $T$ on $\Gamma(t)$ leads to a non-diagonal transformation for $\Gamma(t^\prime)$.
\begin{equation}
\tilde \sigma\Gamma(t)\tilde \sigma^{-1} = \left(\tilde \sigma\Gamma(t \leftarrow t^\prime)\sigma^{-1}\right)\left(\sigma\Gamma(t^\prime)\tilde \sigma^{-1}\right)
\label{divisibility_symmetry_T}
\end{equation}
Inspection of the last term in \eqref{divisibility_symmetry_T} suggests that as long as we want to interpret $\left(\sigma\Gamma(t^\prime)\tilde \sigma^{-1}\right)$ as the transformed $\Gamma(t)$ at time $t^\prime$, compatibility of the left and the right hand sides requires $\sigma=\tilde\sigma$. However, admitting discontinuities, it is no longer a necessity, one can consider $\sigma=\sigma(t)$, $\tilde\sigma=\tilde\sigma$ and one only needs to ensure $\sigma(0)=\tilde\sigma(0)$ and $\sigma(t)=\tilde\sigma(t)$, with arbitrary $\tilde\sigma(t^\prime)$ for $t^\prime\in(0,t)$.

\section{Information decrease}
\label{sec:information}

The diagonal joining the units of the axes in our matrices square corresponds to degenerate $2\times2$ matrices which have identical columns. When $\Gamma(t)$ is on this line, the probability of being at configuration $1$ is the same regardless of starting at configuration $1$ or $2$: any information about the initial probabilities of the system is erased. It is therefore no surprise that the past divisibility cones, the grey area in figures \ref{fig:sequence} representing the region where the curve of figure must be contained in order to be divisible, extends to the entire square once the curve reaches the information-erasing line in the frame \ref{fig:16}.

When $\Gamma(t)$ is at the origin of the square representation ($\Gamma(t)=\sigma_x$) or at $(1,1)=\mathds{1}$, its opposite point, the information about the initial probabilities is fully preserved, since these are the two deterministic and reversible transition matrices. In this sense, the partition of the matrix space according to the principles here explained and exemplified by the grey regions of figures \ref{fig:09_10}, \ref{fig:13} and \ref{fig:sequence} provide a geometric translation of the result of Cohen et al. \cite{Cohen:1993} presented in Buscemi and Datta \cite{Buscemi:2016} that an evolution is divisible if and only the information is monotonically decreasing. We shall now explain this translation in detail. From the previous figures and diagrams it should be clear that the information being decreased is about the original probabilities, since the system may flow to the maximum entropy distribution at the centre of the square as well as to zero entropy distributions corresponding to the edges of the square at $(1,0)$ and $(0,1)$ which represent projectors onto configurations A and B, respectively.

Cohen et al. \cite{Cohen:1993} present a series of entropy inequalities and their connections with ergodicity and contraction coefficients\footnote{Often under the assumption that the stochastic matrix has at least one non-zero element in each row, which may seem restrictive, but the authors of \cite{Cohen:1993} assume generically that the stochastic matrix is not square. One should be in principle careful when the relations require non-empty rows and square stochastic matrices. However, these are often present to avoid divisions by zero, as we shall see in the next section.}. Some of them, although bearing no apparent connection with quantum systems, appear in a curious way when using through the Stochastic-Quantum Correspondence, a result which we derive in section \ref{ssec:ergodicity_contraction}. For now, following \cite{Cohen:1993}, given two probability distributions $\pi$ and $\hat \pi$, define their {\it relative $\phi$-entropy} $H_\phi(\pi,\hat\pi)$.
\begin{equation}
H_\phi(\pi,\hat\pi) = \sum_{i=1}^N \phi(\pi,\hat\pi)
\label{phi-entropy}
\end{equation}

For $2\times2$ matrix, the information inequality associated with \eqref{phi-entropy} is
\begin{equation}
H_\phi(\Gamma\pi,\Gamma\hat\pi) \leqslant |1-(p+q)| H_\phi(\pi,\hat\pi)
\label{lowering-entropy}
\end{equation}
for a stochastic matrix $\Gamma$ such that none of its rows is zero\cite{Cohen:1993}. Note that the proportionality coefficient is nothing but the absolute value of the determinant of $\Gamma$ with its minimum value at the degeneracy line.

Until now we represented stochastic matrices in the square using the probabilities at the diagonal as coordinates. Now we turn our attention to a variant of the decomposition \eqref{Gamma2_deterministic_decomposition} which places the origin at the center of the square, $J$, and has the natural separation of directions: one corresponding to doubly stochastic matrices, between $\mathds{1}$ and $\sigma_x$, and the other connecting the irreversible deterministic matrices, providing the degeneracy direction. This is given by the coordinates
\begin{equation}
X = p-q,~T = 1-(p+q).
\end{equation}
Equivalently, we can write any $2\times2$ stochastic matrix $\Gamma$ as
\begin{equation}
\Gamma
=
X\left(\Pi_A-J\right)+T\left(\sigma_x-J\right)+J
 \label{XTJ}
\end{equation}
with $|X+T|\leqslant 1$ and $|X-T|\leqslant 1$ and
\begin{equation}
J = \frac{1}{4}\left(\mathds{1}+\sigma_x+\Pi_A+\Pi_B\right)
=\frac{1}{2}\begin{pmatrix}1 &1 \\ 1 & 1\end{pmatrix}.
\label{J}
\end{equation}
We shall denote $\Gamma(X,T)$ simply by $(X,T)$.
\begin{equation}
(X,T) = \frac{1}{2}\begin{pmatrix}1+X-T &1+X+T \\ 1-X+T & 1-X-T\end{pmatrix}.
\label{XT}
\end{equation}
The coordinates \eqref{XT} simplify many of the preceding results but, above all, facilitate the physical interpretation of many of them.

The future cone and the associated orientation of information-loss time-direction are all related to the simple expression for the determinant.
\begin{equation}
\det(X,T) = -T
\end{equation}
The future cone is described simply as a convex combination of the current location in matrix space and the line segment $T=0$ for $-1\leqslant X\leqslant 1$ or, in other words, by the set of directions pointing from $\Gamma$ to its divisible future at $T=0$.
The identity is at $T=-1$ and the permutation is at $T=1$. Because a period of divisible evolution implies evolving towards $T=0$, divisibility sets a direction of time {\it in the space of matrices} for divisible dynamics, pointing towards information erasure. The choice of sign for $T$ in \eqref{XT} is the one for which $T$ increases for divisible processes starting at the identity.

Doubly stochastic matrices are located at $X=0$ and these include 2 level systems under unitary time-evolution. Dynamics given, for instance, by $X=p-q=0, T=1-\cos^2(\omega t)$ are divisible until reaching $T=0$ for the first time, evolving after that for a quarter of a period agains the direction of erasure and, therefore, against the information-erasing arrow of time that is associated with divisible dynamics. This is a simple example of how this information-theoretic arrow of time works for Markovian dynamics but may easily fail for quantum dynamics unless we insist on the first and force on the latter a form of retrocausality\footnote{For a review of retrocausal interpretations of quantum theory, see \cite{Friederich:2023}}.

Another advantage of the coordinates \eqref{XT} is the simplification of the multiplication of a two stochastic matrices $(X,T)$ and $(\chi, \tau)$.
\begin{equation}
(X,T)(\chi,\tau) = (X-\chi T, -T\tau)
\label{XT_product}
\end{equation}
Note that at $\tau=0$, action by $(X,T)$ does not take the system out of the degenerate line but shifts it by $X$, so motion along this line is possible even for divisible systems that have lost all initial information. Divisible systems get eventually\footnote{It could be after an infinite amount of parameter time as in a decay process with constant rate $\lambda$.} trapped at $T=0$ but can still move along this $1-dimensional$ subspace of matrix space.

From \eqref{XT_product} it is easy to derive the inverse of a stochastic matrix using the fact that $\mathds{1}=(0, -1)$.
\begin{equation}
(\chi,\tau)^{-1} = (\chi/\tau, 1/\tau)
\label{XT_inverse}
\end{equation}

Equation \eqref{Gamma2_t_t_prime} takes the much simpler form, with $\Gamma(t)=(X,T)$ and $\Gamma(t^\prime)=(\chi,\tau)$.
\begin{align}
\Gamma(t \leftarrow t^\prime)
=
\left(X,T\right)\left(\chi,\tau\right)^{-1} = \left(X-\chi\frac{T}{\tau},-\frac{T}{\tau}\right)
\label{XT_transition}
\end{align}
Inequalities \eqref{n2_inequalities_4} and \eqref{n2_inequalities_6}, defining the past cones, are reduced to the mere requirement that the coordinates in \eqref{XT_transition} have the absolute value of their sum and difference bounded from above by 1.

Another interesting consequence of this parametrisation is that \eqref{XT_transition} easily explanains of the connectivity and the relative density discussed in subsection \ref{ssec:density_connectivity} and illustrated in figures \ref{fig:connectivity_up} to \ref{fig:connectivity_2} in a way that complements the symmetry arguments of section \ref{ssec:symmetries}.

\subsection{Physically questionable cases}

In connection with the information decrease are cases where information increases almost all the time. In practice, our results allow for an easy construction of mathematically sound but physically pathological cases which are indivisible, for example, for all real times except integers. By time-reversal one may obtain curves which are divisible for all real times except for a discrete subset of the continuous time, with isolated instants of indivisibility that only seem indivisible from one ``side" of the parameter time-line.

Consider, for instance, a discrete-time Markov chain with $t\in\mathbb{Z}$ and $\Gamma(t)=\sigma_x^t$. According to the geometric criteria presented here, this is a divisible evolution and for discrete time it could not be otherwise. Now consider the following extension for $t\in\mathbb{R}$.
\begin{equation}
\Gamma(t) =
\begin{cases}
\begin{pmatrix}
2^{-t+2k} & 1-2^{-t+2k}\\
1-2^{-t+2k} & 2^{-t+2k}
\end{pmatrix}, \quad  2k\leqslant t < 2k+1\\
\begin{pmatrix}
1-2^{-t+2k+1} & 2^{-t+2k+1}\\
2^{-t+2k+1}& 1-2^{-t+2k+1}
\end{pmatrix}, \quad 2k+1\leqslant t < 2k+2
\end{cases}
\label{discontinuous_divisible}
\end{equation}
for $k\in \mathbb{Z}$. At time $0$, the evolution starts at the identity and $\Gamma(t)$ goes along the main diagonal towards the centre of the square, the flat matrix, that it would reach at time $t=1$ but it does not, since it jumps to the permutation matrix at $t=1$. From there, $\Gamma$ goes towards the center again along the main diagonal to, at time $t=2$, jump to the identity matrix and repeat the cycle. The curve is right continuous with left limits, or c\`adl\`ag\footnote{``c\`adl\`ag" is a common term, an abbreviation of ``Continue \`a droite, limite \`a gauche", meaning ``continuous on the right, limit on the left".}, continuous, divisible and information-decreasing almost everywhere: information is increased at the discontinuities, leading to instants of indivisibility for $t\in\mathbb{Z}$. Geometrically, the grey zones indicating the past of $\mathds{1}$ for divisibility to hold are two points: $\mathds{1}$ or $\sigma_x$ which clearly does not hold for the real instants immediately before integers.

The time-reverse of \eqref{discontinuous_divisible} is information increasing almost everywhere and, consequently, indivisible for almost every instant, but it is information decreasing when comparing integer times moments with their future neighborhood. In this case, there are instants of divisibility for $t\in\mathbb{Z}$ separated by periods of indivisible evolution for $t\in(k,k+1)$, $k\in\mathbb{Z}$.  This provides an example of an indivisible dynamics with infinitely many events with completely indivisible blocks of time evolution: there exist divisible blocks which are not bi-divisible, in contrast with quantum channels \cite{Davalos:2022wns}. This provides an information-theoretical perspective on how continuity and divisibility are related.

Despite a questionable physical character, it is not excluded that stochastic dynamics for which $\Gamma(t)$ is discontinuous could arise in cases where the dynamics depends on the cumulative distribution function of external discrete variables, which is c\`adl\`ag, or some other analogous external parameters.

\section{Connexions with the Stochastic-Quantum Correspondence}
\label{sec:consequences_stochastic-quantum}

The Stochastic-Quantum Correspondence (SQC) \cite{Barandes:2023ivl,Barandes:2023pwy} is a recent proposal of a many-to-many correspondence between {\it indivisible stochastic dynamics} on one side and {\it quantum} dynamics on the other.
As it fundamentally relies on the indivisibility of the stochastic process and, in particular, on instants where the stochastic evolution is divisible, called division events, our results have consequences for the study of the correspondence. We start in subsection \ref{ssec:stochastic-quantum} with a review of key elements of the SQC, present new connexions\footnote{To the author's knowledge} between the complex-valued description of the SQC and real-valued information and ergodicity measures. In the sequence, subsection \ref{ssec:maximal_maps}, we draw two main consequences: the association of loss of coherences with stochastic divisibility is not bidirectional and the contractive property of stochastic matrices leads to the notion of maximal maps through a geometric counterpart of the validity of the divisibility condition \eqref{divisibility} for all times after the division event.

\subsection{The basics of the correspondence}
\label{ssec:stochastic-quantum}

Every stochastic matrix $\Gamma$ has, by definition, only non-negative entries and, as such, it is always possible to describe each entry of $\Gamma$ as the absolute squared value of a complex number.
\begin{equation}
\Gamma_{ij}^{}(t)=\left|\Theta_{ij}^{}(t)\right|^2
\label{Gamma_to_Theta}
\end{equation}
The entry-wise multiplication of matrices is known as the Schur-Hadamard product and is denoted by $\odot$. With this notation, we can write \eqref{Gamma_to_Theta} as a product of $\Theta(t)$ with itself.
\begin{equation}
\Gamma(t)=\Theta(t)\odot \Theta^*(t)
\label{Gamma_to_Theta_bis}
\end{equation}
Yet another option\footnote{This expression that may inspire retro-causal considerations when $\Theta(t)$ is unitary. In this case, one may interpret $\Gamma(t)$ as the result of a projective measurement over configuration $j$, time evolution from time $0$ to $t$, a projective measurement on configuration $i$, followed by a time-evolution towards the past, or a signal sent to the past, from time $t$ to time $0$. Compare this with the two-state vector formalism \cite{Watanabe:1955,Aharonov:1964}. For a general discussion on retro-causality in Quantum Mechanics see \cite{Friederich:2023}.} is the through the use of the trace and orthogonal projectors
\begin{equation}
\Gamma_{ij}(t)=\text{tr}\left(\Theta^\dagger(t)P_i\Theta(t)P_j\right),
\label{Gamma_to_Theta_3}
\end{equation}
where $P_i$ is an {\it orthogonal} projector onto configuration $i$.

Given initial standalone probabilities $p_i(0)$, the correspondence ascribes the system a density matrix $\rho(t)$ such that
\begin{subequations}
\begin{align}
&\rho(t)=\Theta(t)\rho(0) \Theta^\dagger(t),\\
&\rho(0)=\text{diag}\left({p_1(0), p_2(0), \dots, p_N(0)}\right)
\label{density_matrix}
\end{align}
\end{subequations}
with the associated standalone probabilities at time t, $p_i(t)$.
\begin{equation}
p_i^{}(t)=\text{tr}(P_i^{}\rho(t))
\label{density_matrix_to_probabilities}
\end{equation}
where, again, $P_i^{}$ is an orthogonal projector onto configuration $i$.
Only when there are unitary matrices $\Theta(t)$ that the evolution is unitary, with the corresponding stochastic matrix obtained from \eqref{Gamma_to_Theta} being unistochastic.
Otherwise, the dynamics is that of a general, open quantum system.
The prescription \ref{density_matrix} defines the density matrix and its evolution and it is possible to rephrase it in the language of quantum channels\cite{Barandes:2023ivl,Barandes:2023pwy}. For this define Kraus matrices \cite{Kraus:1971nt} from $\Theta(t)$ by associating each column with a matrix.
\begin{subequations}
\begin{align}
&K_{\alpha}(t):=\vec\Theta_{\alpha}(t)e_\alpha^T,
\label{Theta_to_Kraus_a}
\\
&\sum_{\alpha=1}^N K_\alpha^\dagger(t) K_\alpha(t) = \mathds{1},
\label{Kraus_condition}
\end{align}
\label{Theta_to_Kraus}
\end{subequations}
for all times, where $\vec\Theta_{\alpha}(t)$ is the $\alpha$-th column of $\Theta_{\alpha}(t)$ and $\alpha$ ranges over the columns. Once the matrices are defined via \eqref{Theta_to_Kraus_a}, the Kraus condition \eqref{Kraus_condition} holds by virtue of \eqref{Gamma_to_Theta} and the stochasticity of $\Gamma(t)$. From the Kraus matrices $K_\alpha$, one may rewrite the evolution prescription \eqref{density_matrix} as a dynamical map $\mathcal{E}_t$, i.e., a curve in the space of quantum channels parametrised such that
\begin{equation}
\rho\mapsto\mathcal{E}_t\left(\rho\right):=\sum_{\alpha=1}^{N}K_{\alpha}^{}\left(t\right)X K_{\alpha}^{\dagger}\left(t\right)
\label{quantum_channel}
\end{equation}

Note that the form \ref{Theta_to_Kraus_a} of the Kraus matrices prescribed in equation \ref{quantum_channel} following \cite{Barandes:2023pwy} are such that all initial coherences of $\rho$ are erased, since only \eqref{quantum_channel} acts only on the diagonal of $\rho$. In the case $N=2$ they correspond, at a given time $t$, to divisible a quantum channel, since indivisible channels have Kraus rank 3 \cite{Davalos:2018,Ende:2024ksq}. This does not mean that the dynamical map is divisible. For a computational approach towards channel divisibility see \cite{Nery:2024fyl}. We insist that the overall approach in this paper does not place a restriction on the Kraus rank because one may start with a quantum system and derive a stochastic representation of the dynamics as presented in the introduction, equation \ref{channel_to_Gamma}.

When $\Gamma(t)$ is not unistochastic, it is possible to dilate the Hilbert space description using Stinespring's dilation theorem \cite{Stinespring:1955} and to show that every indivisible stochastic dynamics can be described as the dynamics of a subsystem where the larger system evolves unitarily. This is the content of the Stochastic-Quantum Theorem \cite{Barandes:2023pwy}: every generalised stochastic system can be seen as a subsystem of a quantum system evolving unitarily in a Hilbert space.

For the derivation of other aspects of quantum theory from indivisible stochastic dynamics, such as Hamiltonian evolution, dilations, gauge symmetries, the reader is referred to \cite{Barandes:2023ivl,Barandes:2023pwy}.

The reader should keep in mind that (1) all the results here obtained are translatable to the language of quantum channels but the concept of divisibility used here is of stochastic divisibility, meaning that application of our results is more convenient when translating quantum channels to stochastic matrices and applying our divisibility criteria; (2) we emphasise that relying solely on probabilities does not exclude quantum phenomena when {\it indivisivible dynamics} is taken into account and; (3) the possibility of choosing freely the standalone probabilities $p_i(0)$ and $\Gamma(0)=\mathds{1}$ is associated with the orthogonality at $t=0$ of the corresponding quantum states but it follows from \eqref{Theta_to_Kraus} and \eqref{density_matrix} that initially orthogonal states may evolve to non-orthogonal states as a stochastic matrix $\Gamma$ is a convex combination of permutations {\it and} projectors.

\subsection{Contraction, ergodicity and information again}
\label{ssec:ergodicity_contraction}

In section \ref{sec:information} we briefly discussed the relative $\phi$-entropy and showed how they provide a natural time in the space of stochastic matrices for divisible processes. In this section we discuss how other information inequalities presented in \cite{Cohen:1993} have their contraction or ergodicity coefficients related to a measure of non-unitarity that we propose based on the STC construction \cite{Barandes:2023ivl}.

The matrices $\Theta$ obtained form $\Gamma$ by the requirement \eqref{Gamma_to_Theta} do not need to be unitary and, in general, are not. We shall call $\Theta$ a {\it Schur-Hadamard square-root} of $\Gamma$, due to equation \ref{Gamma_to_Theta_bis}, or simply a {\it SH square-root} of$\Gamma$. The problem of identifying unistochastic matrices, i.e., stochastic matrices which admit a choice of phases that renders their SH square-root unitary is analytically known only for $N=2,3$ \cite{Fedullo:1992} with interesting algebraic and geometric structures for N=3 \cite{Jarlskog:1985ht,Rajchel-Mieldzioc:2021ohk}.

In the $2\times2$ case, with $(p,q)$ parametrisation \eqref{Gamma2}, a necessary and sufficient condition for the unistochasticity is the equality of the diagonal entries $p$ and $q$. Once this is satisfied, it is necessary to appropriately chose the phases of the SH square-root. An interesting thing happens when one considers matrices $\Theta$ which are not unitary\footnote{We could have said special unitary without any loss of generality but this option leads to a more symmetric expression for Birkhoff's contraction coefficient in terms of $\Theta$.}  but choses phases that would render $\Theta$ unitary once equality of $p$ and $q$ holds.
\begin{equation}
\Theta=e^{i\phi}
  \begin{pmatrix}
  \sqrt{p} & -\sqrt{1-q}e^{i\theta}\\
  \sqrt{1-p}e^{-i\theta} & \sqrt{q}
  \end{pmatrix}
\label{quasi_unitary}
\end{equation}
With this choice, we can see the absolute value of the determinant of $\Theta$ as a measure of the deviations from unitarity.
\begin{equation}
|\det\Theta |=\sqrt{pq} +\sqrt{(1-p)(1-q)}
\label{unitarity_deviation}
\end{equation}
Since $\Theta$ is a SH square-root of $\Gamma$, it is also interesting to compare the ratios of the determinant of $\Gamma$ to the square of the determinant of $\Theta$.
\begin{equation}
\frac{\det\left(\Theta\odot \Theta^*\right)}{\det\left(\Theta\Theta*\right)}=\frac{1-(p-q)}{\left[\sqrt{pq} +\sqrt{(1-p)(1-q)}\right]^2}=-\tau_B(\Gamma)
\label{Birkhoff_Theta}
\end{equation}
The $\tau_B(\Gamma)$ in \eqref{Birkhoff_Theta} is Birkhoff's contraction coefficient of $2\times2$ stochastic matrix with diagonal $(p, q)$ excluding the deterministic irreversible cases $(0,1)$ and $(1,0)$.

Expression \ref{Birkhoff_Theta} suggests a connection between the complex description from the SQC to a purely real result for Birkhoff's contraction coefficient $\tau_B(\Gamma)$ which is defined, in general, by
\begin{equation}
\tau_B(\Gamma)=\sup\Set{\frac{d\left(\Gamma x, \Gamma y\right)}{d\left(x, y\right)} ~:~x, y \in \mathbb{P}_N,~x\neq y}
\label{Birkhoff}
\end{equation}
where $\mathbb{P}_N$ is the space of probability distributions on $N$ configurations, $d$ is the Hilbert projective pseudo-metric,
\begin{equation}
d(x,y) = \max_{i,j}\log\frac{x_iy_j}{x_jy_i}
\label{Hilbert_pseudometric}
\end{equation}
for $x, y$ positive $N$-vectors. The distance \eqref{Hilbert_pseudometric} is an upper bound for the symmetric relative entropy $J(x,y)=H(x,y)+H(y,x)$ where $H(x,y)=\sum_i x_i\log(x_i/y_i)$ and this last one vanishes iff $J=d$ and iff both positive vectors are equal \cite{Cohen:1993}.

Finally, the determinant of $\Gamma$ appearing in equation \ref{lowering-entropy}, another information inequality, is called Dobrushin's coefficient of ergodicity and it coincides for $N=2$ with Doeblin's coefficient of ergodicity \cite{Cohen:1993}. The supremum of the ratio of $H_\phi(\Gamma\pi,\Gamma\hat\pi)$ and $H_\phi(\pi,\hat\pi)$ introduced in section \ref{sec:information} is a measure of scrambling which can also be expressed, for $N=2$ in terms of $\det(\Gamma)$ and $\det(\Theta)$, as follows from \eqref{quasi_unitary} and table 1 of \cite{Cohen:1993}. Finally, with a $\hat\Theta$ defined by a choice of phases that reverses the sign in \eqref{quasi_unitary} one can also write $\det\Gamma$ as $\det(\Theta\hat\Theta)$. Which of these coincidences, if any, remains valid for $N>2$ as well as the physical interpretation of these relations from the point of view of the complex representation or any of its Hilbert space dilations are subjects for future investigations.

\subsection{Maximal maps as selectors of division events}
\label{ssec:maximal_maps}

Associating indivisible stochastic dynamics with quantum dynamics is the key feature of the Stochastic-Quantum Correspondence.
The above analysis, even dealing with the simplest non-trivial system, implies that not all division events are on the same footing. Up to know we considered $\Gamma(t)$ at a given time $t$ and analysed its past and future. This is summarised in figure \ref{fig:11_12} illustrates the effect of acting on $\Gamma(t)$ with all possible $2\times2$ stochastic matrices, the yellow parallelogram at the center, and the effect of acting on any stochastic matrix of the same dimension with $\Gamma(t)$, the magenta square at the center. In both cases, regardless of the (in)divisibility of what follows or precedes, the space of matrices possible is considerably reduced. There are two possible scenarios: either the evolution is faithfully captured by the division events at time $t$ and time $0$ or describing the dynamics by dividing $\Gamma(t)$ will not cover the entire evolution. In the second case, one may look for an extended dynamics $\Gamma_E(t)$ which covers the entire evolution.

Again, the oscillator example is very useful to illustrate this point. The process as a whole is indivisible but the dynamics is divisible as long as the stochastic matrix moves towards the centre.
\begin{equation}
\Gamma(t)
=
\begin{pmatrix}
  \cos^{2}{\left(\pi t /2\right)} & \sin^{2}{\left(\pi t /2\right)}\\
  \sin^{2}{\left(\pi t/2 \right)} & \cos^{2}{\left(\pi t/2 \right)}
\end{pmatrix}
\label{Gamma_cos_t}
\end{equation}
There are infinitely many values of $t^\prime$ and such that \eqref{divisibility} holds for \eqref{Gamma_cos_t}. Some of them are given by $0\leqslant t^\prime\leqslant t< 1$.
\begin{equation}
\Gamma(t\leftarrow t^\prime)
= \frac{1}{2} +
  \frac{\cos\left(\pi t \right)}{2\cos\left(\pi t^\prime \right)}
\begin{pmatrix}
  1 &-1\\
  -1 & 1
\end{pmatrix}
\label{Gamma_cos_ttp}
\end{equation}
The transition matrix \eqref{Gamma_cos_ttp} is not stochastic for every $t$. When $t$ goes slightly below $s$, the entries of $\Gamma(t\leftarrow s)$ are above 1 or below 0. Also, $t$ in $\Gamma_ts$ does not extend arbitrarily far into the future as in \eqref{Gamma_cos_t} with the same issue of coefficients above 1 or below 0 arising. As $t'$ approaches 1, the range over which one can continuously extend \eqref{Gamma_cos_ttp} for $t>1$ is reduced and goes to zero as $t^\prime$ approaches $1$. This is a manifestation of the phenomenon captured by the yellow and magenta zones of figure \ref{fig:11_12} and motivates the following definitions.

A stochastic dynamical map given by a curve $\Gamma(t)$ in the space of stochastic matrices is called {\bf maximal} if it covers all the possible values of $t$ associated with the underlying stochastic system.
A division event at time $t^\prime$ is considered {\bf proper} if the resulting map $\Gamma(t\leftarrow t^\prime)$ is maximal.

From these definitions it follows that every passage through a permutation matrix constitutes a proper division event. However, if the system naturally evolves to a reduced area of the space of stochastic matrices, a proper division event is allowed to occur at $t^\prime>0$, with a maximal map that contracts the space of matrices compared to the one at parameter time $0$.

Matrix \ref{Gamma_cos_t} being unistochastic, we can apply to it the prescription \ref{density_matrix} to obtain a unitary time-evolution for the density matrix with division events happening for $t\in[0,1)$ while $\rho(t)$ has non-zero coherences. However, {\it proper} division events happen for times $t$ when  \ref{Gamma_cos_t} correspond to permutation matrices and the coherences from prescription  \ref{density_matrix} vanish.

The distinction between division events in the past of $\Gamma(t)$, given a specific time $t$ and proper division events requires knowledge (or prescription) of the full future of the stochastic evolution after a specific time $t$ and has a teleological aspect, as the event horizon of a black hole does.

\section{Coarse graining and dilations}
\label{sec:coarse}

Open quantum systems under non-unitary evolution can be enlarged to a dilated Hilbert space where the evolution is unitary, as proven by Stinespring \cite{Stinespring:1955}. An analogous result for stochastic matrices was proven by Schmidt \cite{Schmidt:2021} with two different types of dilation: a stochastic environmental dilation and a dilation through coarse graining. The environmental dilations can be seen in two equivalent ways: first, by representing $\Gamma$ as a fully decohering quantum channel that acts only on probabilities, dilating it using Stinespring's theorem and converting the channel back to a stochastic matrix; second by directly finding a larger stochastic matrix such that, for a specific configuration of the auxiliary system, marginalising over the auxiliary system reproduces the original stochastic matrix. Schmidt showed that these are both equivalent to {\it dilations by coarse-graining} which mapping stochastic matrices to bistochastic ones in the dilated space. For this reason we work directly with dilation by coarse graining. In this section we present the procedure through examples, generalise the coarse-graining procedure by including uncertainty and the dilation by coarse-graining by including dynamics. Applications to the study of indivisible stochastic dynamics, clarify some discussed above and connects the case of 2 configurations with more general cases.

Let $\mathcal{L}$ be a (large) system with $N$ configurations and $\mathcal{S}$ be a (smaller) system with configurations labeled by $n<N$ configurations. {\it Coarse graining} and {\it dilation by coarse graining} are achieved by
\begin{enumerate}[label=(\roman*)]
\item an $n\times N$ deterministic, left-stochastic matrix X that projects probability distributions over $\mathcal{L}$ onto probability distributions over $\mathcal{S}$ and
\item a $N\times n$ left-stochastic matrix Y that is (non-unique) right-inverse of X.
\end{enumerate}

We illustrate the procedure through examples and show how to introduce time-dependence in a consistent way.

\subsection{Coarse graining stochastic dynamics}

Consider a system $\mathcal{L}$ with $N=6$ configurations and $\mathcal{S}$ with $n=2$ configurations. One possible way to coarse grain the system is represented by the following $2\times6$ stochastic matrix
\begin{equation}
X
=
\begin{pmatrix}
1 & 1 & 1 & 1 & 0 & 0 \\
0 & 0 & 0 & 0 & 1 & 1
\end{pmatrix}
\label{X42}
\end{equation}
Being left-stochastic, matrix $X$ maps probability distributions over $\mathcal{L}$ to probability distributions over $\mathcal{S}$.

If probabilities in $\mathcal{L}$ evolves according to
\begin{equation}
p^{(\mathcal{L})}_{I}(t)=\sum_{J=1}^N\Gamma^{(\mathcal{L})}_{IJ}(t)p^{(\mathcal{L})}_{J}(0)
\label{pLarge}
\end{equation}
then, the evolution of probabilities in $\mathcal{S}$ evolve as
\begin{equation}
p^{(\mathcal{S})}_{i}(t)=\sum_{J=1}^N X^{}_{iJ}p^{(\mathcal{L})}_{IJ}(t)\sum_{J, K=1}^N X_{iJ}\Gamma^{(\mathcal{L})}_{JK}(t)p^{(\mathcal{L})}_{K}(0).
\label{pSmall}
\end{equation}

Equation \eqref{pSmall} is general. The product $X\Gamma^{(\mathcal{L})}$ maps $p^{(\mathcal{L})}(0)$ to $p^{(\mathcal{S})}(t)$, which is desirable but not sufficient if one wants to fully describe the evolution in terms of the coarse grained system only: that is where a right-inverse is necessary.

Let $Y(0)$ be a left-stochastic matrix that is a right-inverse\footnote{A right-inverse is guaranteed to exist from the fact that $X$ represents a {\it proper} coarse graining, implying that $X$ is surjective and has full row rank. A right-inverse being non-unique, there is, at this point, no reason to assume one should choose the same at different times.} of $X$, i.e. $XY(0) = \mathds{1}_n$. Then we can define $\Gamma^{(\mathcal{S},Y)}(t)$ such that
\begin{equation}
p^{(\mathcal{S})}_{i}(t)=\sum_{j=1}^n\Gamma^{(\mathcal{S},Y)}_{ij}(t)p^{(\mathcal{S})}_{j}(0)
\label{pGammaSmall}
\end{equation}
via
\begin{equation}
\Gamma^{(\mathcal{S}, Y)}_{ij}(t) = \sum_{I,J=1}^N X_{iI}\Gamma^{(\mathcal{L})}_{IJ}(t)Y_{Jj}(0).
\label{GammaSmall_ij}
\end{equation}
or, compactly,
\begin{equation}
\Gamma^{(\mathcal{S}, Y)}(t) = X\Gamma^{(\mathcal{L})}(t)Y(0).
\label{GammaSmall}
\end{equation}

Because $Y(0)$ is non-unique, we keep a $Y$ superscript as a reminder of the fact that different choices of $Y(0)$ will lead to different dynamics for the underlying system.

\begin{figure}[!htp]
\centering
\includegraphics[width=0.45\textwidth]{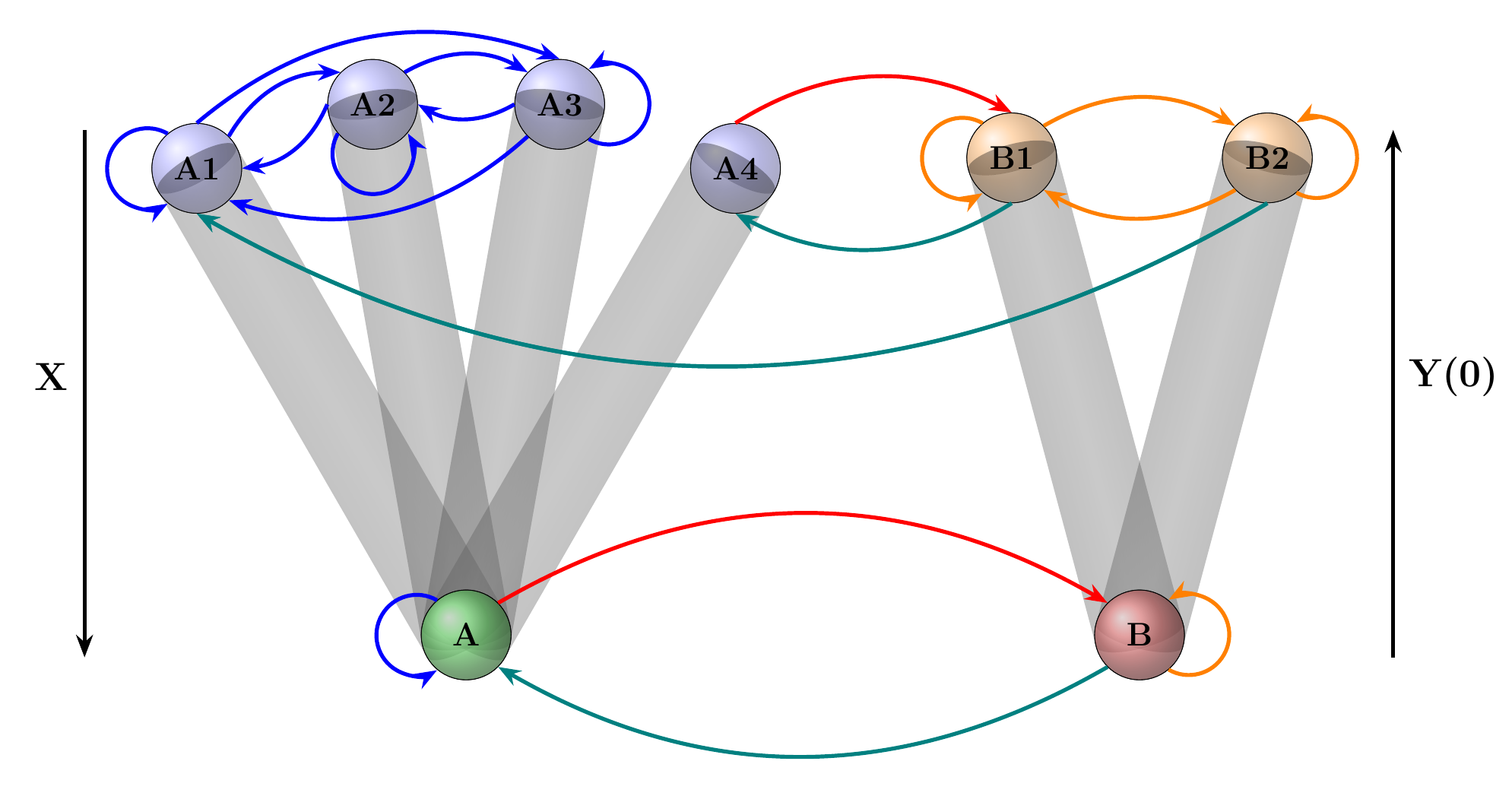}
\caption{An example of a coarse graining by grouping the nodes $A1$, $A2$, $A3$ and $A4$ as node $A$ and nodes $B1$ and $B2$ as node $B$. The arrows represent non-vanishing coefficients of a stochastic matrices, $\Gamma^{(\mathcal{L})}(t)$ defined on the larger system and $\Gamma^{(\mathcal{S},Y)}(t)$ defined on the smaller system by equation \eqref{GammaSmall}. The presence of the red arrow on the coarse-grained system depends upon the choice of $Y(0)$}
\label{fig:24}
\end{figure}

At this point, the use of $Y(0)$ instead of $Y$ is optional but as soon as we introduce dilations, we will consider curves $Y(t)$ in the space of $N\times n$ left-stochastic matrices that are right-inverses of $X$. We postpone until then the discussion of the physical meaning of the time-dependence of $Y$. Now we limit ourselves to discuss the physical meaning of the $Y$ dependence in $\Gamma^{(\mathcal{S},Y)}(t)$. For this, we continue with the example of matrix \eqref{X42} and apply the procedure to a pair $\Gamma^{(\mathcal{L})}(t)$, $\Gamma^{(\mathcal{S},Y)}(t)$ represented in figure \ref{fig:24} through a pair of graphs.

The possible matrices $Y(0)$ are:
\begin{equation}
Y(t)
=
\begin{pmatrix}
Y_{A_1A}(0) & 0\\
Y_{A_2A}(0) & 0\\
Y_{A_3A}(0) & 0\\
Y_{A_4A}(0) & 0\\
0 & Y_{B_1B}(0)\\
0 & Y_{B_2B}(0)
\end{pmatrix}
\label{Y42}
\end{equation}
with $Y_{Ij}\geqslant 0$ and $\sum_{I=A_1}^{B_2}Y_{Ii}=1$ for $i=A, B$.

All other possible coarse grainings that partition the system in a block of four configurations and one of two are related to the matrix \eqref{X42} and and the set of matrices \eqref{Y42} by permutations\footnote{More generally, given a partition of $\mathcal{L}$ and a surjective map from this partition to $\mathcal{S}$ via a deterministic stochastic matrix X, all other partitions of $\mathcal{L}$ with the same partition sizes can be obtained from X via $\sigma^{(\mathcal{S})}_i$X$\sigma^{(\mathcal{L})}_J$, where $\sigma^{(\mathcal{S})}_i$,$i=1,\dots,n!$ is a permutation matrix in $\mathcal{S}$, and  $J=1,\dots,N!$. As a right inverse, $Y$ should be mapped to $\left(\sigma^{(\mathcal{L})}_J\right)^{-1}$Y$\left(\sigma^{(\mathcal{S})}_i\right)^{-1}$}.

Assuming that the connectivity of figure \ref{fig:24} is that of time $t$, this implies that $\Gamma^{(\mathcal{L})}_{I,A_4}(t)=\delta_{I,B_1}$ and, consequently, that upon coarse graining through \eqref{GammaSmall} the conditional probability of finding the system in configuration B at time $t$ given that it was in A at time $0$ is
\begin{equation}
\Gamma^{(\mathcal{S},Y)}_{BA}(t) = Y_{A_4A}(0)
\end{equation}

A different choice of coarse graining, grouping $A_1$, $A_2$ and $A_3$ as $\tilde{A}$ and $A_4$, $B_1$ and $B_2$ as $\tilde{B}$ would avoid this issue in the present example for time $t$. However, since indivisible stochastic dynamics allows for arbitrary changes of connectivity over time, the problem could reappear for different times, depending on the dynamics, leading to a $Y(0)$-dependent connectivity. Physical principles may impose constraints on connectivity changes and it is conceivable that, for certain problems, a good choice of coarse graining should minimise this dependence.

\begin{figure}[t]
\centering
\includegraphics[width=0.49\textwidth]{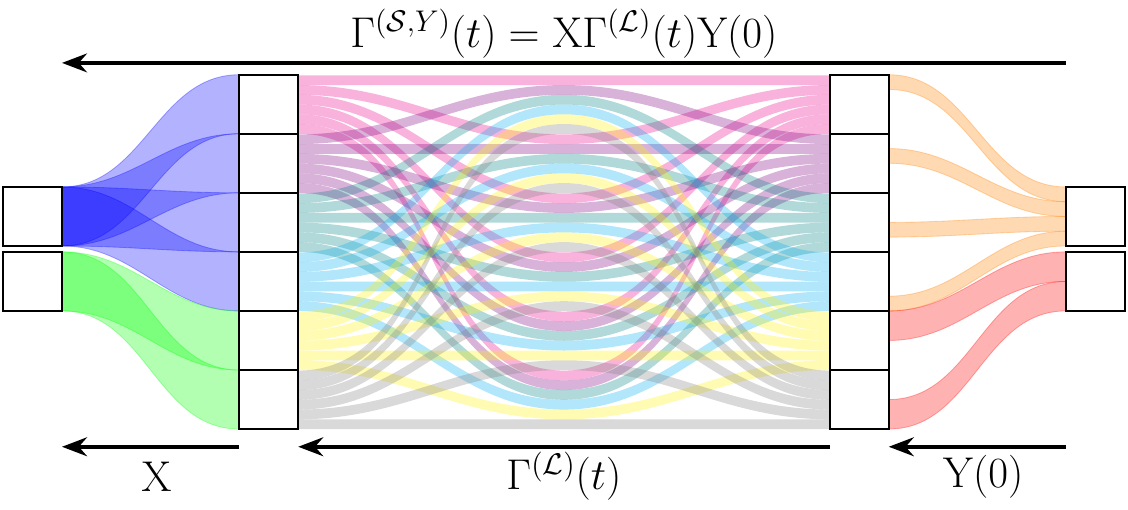}
\caption{A dynamical perspective of the coarse graining by grouping the nodes $A1$, $A2$, $A3$ and $A4$ as node $A$ and nodes $B1$ and $B2$ as node $B$. Matrix $Y(0)$ dispatches a coarse grain into thin grains, the probabilities evolve according to $\Gamma^{(\mathcal{L})}_{}(t)$ and matrix $X$ performs a final grouping of configurations to produce $\Gamma^{(\mathcal{S},Y)}(t)$ according to \eqref{GammaSmall}.}
\label{fig:squares_grow_mix}
\end{figure}

Figure \ref{fig:squares_grow_mix} summarises the procedure, with a different time evolution and uniform coarse graining, i.e., each non-zero value of a column of $Y(0)$ has the same value.
Different choices of $Y(0)$ will in general lead to different initial conditions for the larger system. The probabilities evolve according to $\Gamma^{(\mathcal{L})}_{}(t)$ and matrix $X$ performs a grouping of configurations. The resulting transformation $\Gamma^{(\mathcal{S},Y)}(t)$ maps probabilities on $\mathcal{S}$ to probabilities on $\mathcal{S}$ and is a proper coarse-grained stochastic matrix.

For a more detailed treatment of the uniform case and a definition of the general form of $X$ in terms of indicator functions, see \cite{Schmidt:2021}.

\subsection{Dynamics and dilations by coarse graining}
\label{ssec:dilation_coarse}

Dilation by coarse graining is simply defined by reversing the order in the above procedure.
Given a system $\mathcal{S}$ with $n$ configurations, chose an arbitrary larger system $\mathcal{L}$. It is possible to dilate $\mathcal{S}$ to $\mathcal{L}$ by first coarse-graining the probabilities from $\mathcal{L}$ to $\mathcal{S}$ with a deterministic left-stochastic matrix $X$ with no empty rows and evolving the probabilities to, finally, dilate them with a right-inverse of $X$:
\begin{equation}
\Gamma^{(\mathcal{L}, Y)}_{IJ}(t) = \sum_{I,J=1}^N Y_{Ii}(t)\Gamma^{(\mathcal{S})}_{ij}(t)X_{jJ}.
\label{GammaBig_IJ}
\end{equation}
or, compactly,
\begin{equation}
\Gamma^{(\mathcal{L}, Y)}(t) = Y(t)\Gamma^{(\mathcal{S})}(t)X.
\label{GammaBig}
\end{equation}
where the same rules of indices apply. Probability distributions in the large system evolve according to
\begin{equation}
p^{(\mathcal{L}, Y)}(t)=\Gamma^{(\mathcal{L}, Y)}(t)p^{(\mathcal{L})}(0) = Y(t)\Gamma^{(\mathcal{S})}(t)\left[Xp^{(\mathcal{L})}(0)\right].
\label{dilation_pLY}
\end{equation}

\begin{figure}[t]
\centering
\includegraphics[width=0.43\textwidth]{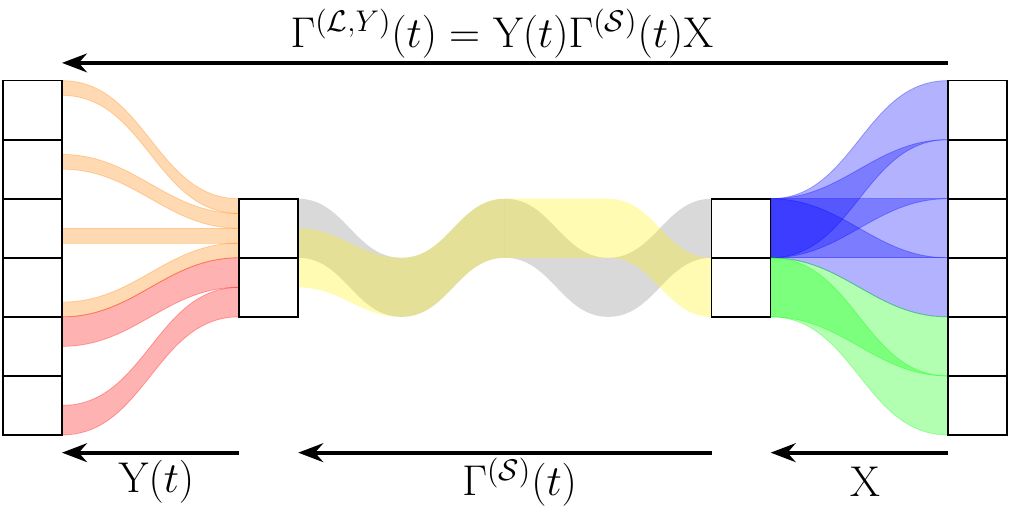}
\caption{A dilation by coarse graining of a system $\mathcal{S}$ with dynamics ruled by $\Gamma^{(\mathcal{S})}_{}(t)$. $X$ merges configurations while $Y(t)$ splits them with $t$-dependent proportions (probabilities).
The resulting transformation $\Gamma^{(\mathcal{L},Y)}(t)$ maps probabilities on $\mathcal{L}$ to probabilities on $\mathcal{L}$ and is a proper dilation by coarse-grained of the stochastic dynamics.}
\label{fig:26}
\end{figure}

Figure \ref {fig:26} illutrates this principle.

The initial conditions in \eqref{dilation_pLY} render the probabilities for $t\neq0$ insensitive to the choice of $Y(0)$, since the initial conditions are given by $\left[Xp^{(\mathcal{L})}(0)\right]$, the coarse-grained initial probability distribution\footnote{Given an initial probability distribution over $\mathcal{S}$ at time $0$, two different dilations will correspond to $p^{(\mathcal{L}, Y^{(1)})}(0)=Y^{(1)}(0)p^{(\mathcal{S})}(0)$  and $p^{(\mathcal{L}, Y^{(2)})}(0)=Y^{(2)}(0)p^{(\mathcal{S})}(0)$ which evolve according to \eqref{dilation_pLY} but $XY^{(1)}=XY^{(2)}=\mathds{1}_{\mathcal{S}}$. Similarly, any two distributions over $\mathcal{L}$ that lead to the same distribution over $\mathcal{S}$ will lead to the same probabilities under the time evolution described by \eqref{dilation_pLY}.
}.

Now, time-evolution in $\mathcal{L}$ depends on the choice of a curve $Y(t)$ of left-stochastic right-inverses of $X$. The matrix $Y(t)$ at a given time may represent an expected underlying distribution on real, but otherwise not well-known, subdivisions of the system.

Indivisible dynamics prevents conditioning at times of indivisible evolution and, as such, it is natural not to require $Y(t)$ to be time-translation invariant. Things could change regarding divisibility but, in fact, when the dynamics in $\mathcal{S}$ is divisible at time $t$, so is the dilated dynamics, for
that divisibility of the a system implies divisibility of its dilated version since
\begin{equation}
Y(t)\Gamma(t)X = Y(t) \Gamma(t \leftarrow t^\prime)XY(t^\prime)\Gamma(t^\prime)X.
\label{divisibility_dilated1}
\end{equation}
and $XY(t^\prime)=\mathds{1}_n$.

\begin{figure}[!htp]
\centering
\begin{minipage}{0.23\textwidth}
\subfloat[XY=$\mathds{1}_n$.]{
\includegraphics[width=\textwidth]{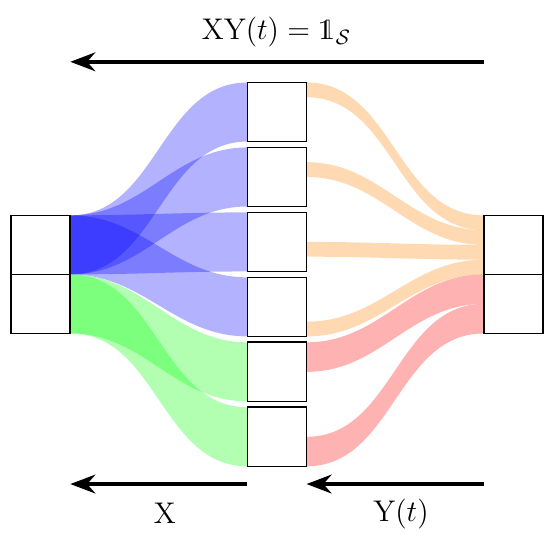}
\label{fig:27}
}
\end{minipage}
\begin{minipage}{0.23\textwidth}
\subfloat[YX is always degenerate.]{
\includegraphics[width=\textwidth]{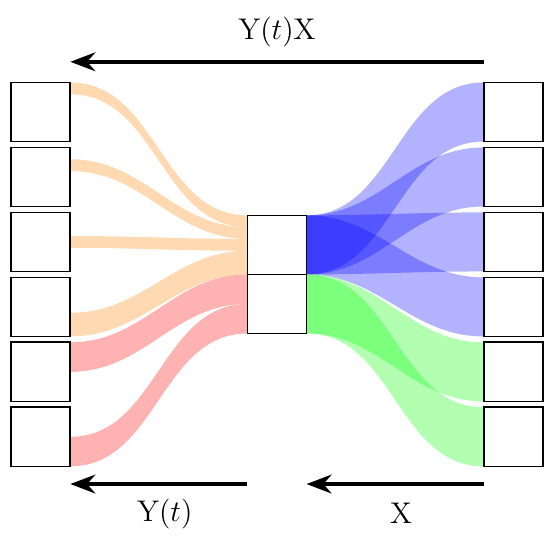}
\label{fig:28}
}
\end{minipage}
\caption{Comparison of $XY$ and $YX$ where $X$ is the grouping matrix, and $Y$ any of its right-inverses. Figures \ref{fig:27} and \ref{fig:28} correspond to different matrices $Y$.}
\label{fig:27_28}
\end{figure}

Finally, when the stochastic matrix being coarse-grained or dilated is the identity, we obtain a graphical depiction of $Y$ as a right-inverse of $X$ and of $XY$ as a left-stochastic matrix with duplicated columns, as shown in figure \ref{fig:27_28}.

Figure \ref{fig:28} illustrates equation \eqref{dilation_pLY} for $\Gamma=\mathds{1}_n$ in the sense that for different choices of initial configuration $j$, the probabilities of arrival at configuration $i$ will be the same, thus producing duplicated columns.

In section \ref{ssec:divisibility_coarse} we will discuss the non-trivial question of how divisibility of the dynamics of a large system relates to the divisibility of the coarse-grained dynamics.

\subsection{Degeneracy}
\label{ssec:dilation_degeneracy}

Every dilation by proper coarse-graining produces a degenerate stochastic matrix\footnote{The column indices in \eqref{GammaBig_IJ} are those of $X$ and for a ``proper" coarse-graining, as Schmidt puts it \cite{Schmidt:2021} requires that $X$ has more columns than rows. Being deterministic, $X$ will have at least two identical columns and so will any dilated matrix obtained by \eqref{GammaBig_IJ}.}. But not every degenerate stochastic matrix is a dilation by coarse graining, except when the dilated system has $2$ configurations.

\begin{figure}[!htp]
\centering
\includegraphics[width=0.25\textwidth]{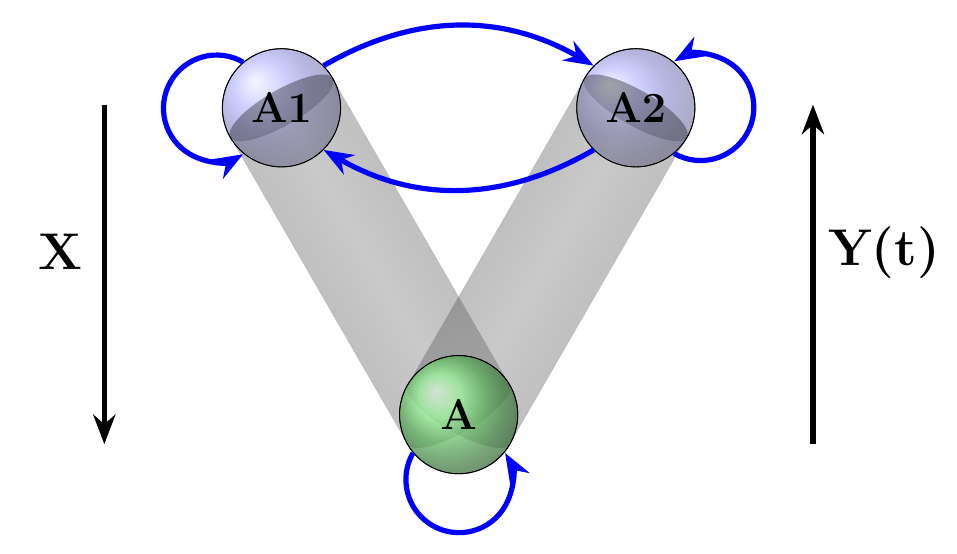}
\caption{An example of a coarse graining by grouping the nodes $A1$, $A2$ as node $A$ through the matrix $X$. Although the coarse graining produces a trivial dynamics, dilation with time-dependent $Y(t)$ produces all degenerate $2\times2$ dynamics.}
\label{fig:a12}
\end{figure}

This means that the secondary diagonal of figures \ref{fig:09_10}- \ref{fig:sequence}, which correspond to matrices of the form \eqref{Gamma2_degenerate} with associated graphs \ref{fig:03_04_05}, are the result of a dilation by coarse graining of a the trivial stochastic system with one configuration. The corresponding matrix $X$ is
\begin{equation}
X
=
\begin{pmatrix}
1 & 1
\end{pmatrix}
\end{equation}
and the allowed right inverses of $X$ are probability distributions on two variables. We write
\begin{equation}
Y(t)
=
\begin{pmatrix}
y(t) \\ 1-y(t)
\end{pmatrix},\quad y(t)\in[0,1].
\end{equation}
The single node has trivial dynamics since its only entry $\Gamma_{AA}(t)$ must equal one for all times. For this reason, the dilated system has dynamics given by
\begin{equation}
\Gamma^{(dilated)}(t)
=
Y(t)X
=
\begin{pmatrix}
y(t) & y(t)\\ 1-y(t) &1-y(t)
\end{pmatrix}.
\end{equation}
Figure \ref{fig:a12} represents this procedure of mapping 2 configurations to 1 via X and the action of X's right inverse $Y(t)$ mapping 1 to 2 configurations.

In fact, all stochastic matrices obtained via dilations by dynamical coarse graining are degenerate, but not all degenerate matrices are equivalent to dilations by coarse graining\footnote{The reason is quite simple. For X to constitute a proper coarse graining, it must be a deterministic $n\times N$ left-stochastic matrix with $N>n$. As such, it must contain at least two identical columns. The components of $Y\Gamma X$ are $\sum_{k=1}^n\sum_{k=1}^nY_{ik}\Gamma_{k\ell}X_{\ell j}$ where $n$ is number of configurations of the non-dilated system. Therefore, at least two columns of $Y\Gamma X$ are identical, proving its degeneracy. However, for $N\geqslant3$ it is possible to have a zero determinant without identical columns. The requirement of a ``proper" coarse graining, i.e. $X$ has no empty rows, does not change this result.
}.

Before moving forwards, a quick note on {\it improper} coarse-graining, a straightforward generalisation Schmidt's approach of dilation by coarse-graining to allow for grouping under uncertainty or to represent groupings done by some random procedure. This is effectuated by considering left-stochastic $n\times N$ matrices $X$ which are {\it not} deterministic. The main difference from the {\it proper} coarse-graining is that $X$ has no longer a left-stochastic right-inverse.  To illustrate the principle, consider a system with 3 configurations $\mathcal{C_L^{}}=\{1,2,3\}$ and another with 2 configurations, $\mathcal{C_S^{}}=\{A,B\}$. Consider a mapping $X$ that sends configuration $1$ to $A$, configuration $3$ to $B$ but that maps configuration $2$ to $A$ with probability $p$ and to $2$ to $B$ with probability $1-p$. The grouping matrix is
\begin{equation}
X
=
\begin{pmatrix}
1 & p & 0\\
0 & 1-p & 1
\end{pmatrix}=
p
\begin{pmatrix}
1 & 1& 0\\
0 & 0 & 1
\end{pmatrix}
+
(1-p)
\begin{pmatrix}
1 & 0 & 0\\
0 & 1 & 1
\end{pmatrix}.
\end{equation}
Given its physical meaning of uncertainty in the grouping, either by ignorance in the modelling or by randomness in the process, it is natural that the general procedure consists in decomposing $X$ in a left-stochastic deterministic basis and by mixing the dilations by a convex combination with the coefficients of the decomposition. A $2\times 2$ matrix $\Gamma$ is dilated according to
\begin{equation}
\Gamma^{(\mathcal{L})}= p Y_1\Gamma X_1+ (1-p) Y_2\Gamma X_2.
\end{equation}
In the general case, given a left-stochastic $n\times N$ matrix X, we write
\begin{equation}
X = \sum_{k=0}^{nN}\alpha_k X_k, \quad \sum_{k=0}^{nN}\alpha_k = 1,
\label{X_uncertainty}
\end{equation}
and choose left-stochastic $Y_k$, $k=1, 2, \dots, nN$ such that $X_kY_k=\mathds{1}_n$. This is closer in spirit to the approach of \cite{Duarte:2017} for coarse-graining quantum systems but instead of a dilation via Stinespring's theorem \cite{Stinespring:1955}, we modify Schmidt's $XY$ procedure as follows. A coarse graining under uncertainty of an $N\times N$ stochastic matrix $\Gamma$ is defined as the matrix
\begin{equation}
\Gamma^{CG} := \sum_{k=0}^{nN}\alpha_k X_k \Gamma Y_k
\label{CG_uncertainty}
\end{equation}
and a dilation by coarse-graining under uncertainty of an $n\times n$ stochastic matrix $\hat \Gamma$ is defined by
\begin{equation}
\Gamma^{DCG} := \sum_{k=0}^{nN}\alpha_k Y_k \Gamma X_k.
\label{DCG_uncertainty}
\end{equation}
Each $Y_k$ represents a dispatching procedure that sends the elements from coarse-grained configurations back to (at least part of) the original slots in the large system, with a specific $(X_k, Y_k)$ pair represented in each of figures \ref{fig:26} and \ref{fig:27_28}, and randomly this procedure with probability $\alpha_k$. 
Another motivation for this generalisation is to generate, from lower-dimensional systems, sets of degenerate matrices in larger spaces. One should keep in mind, however, that a dilation by coarse graining under uncertainty does not necessarily preserve divisibility of the dynamics, although a dilation by proper coarse-graining does. That said, now we investigate under which circumstances a coarse-graining preserves divisibility.

\subsection{Divisibility and coarse graining}
\label{ssec:divisibility_coarse}

In subsection \ref{ssec:dilation_coarse} we saw that {\it dilations} by proper coarse-graining preserve divisibility.
In this section we show that (in)divisibility of the evolution at a given time is not preserved under coarse-graining.

For this, let $\mathcal{C}_{O}=\{1,2,\dots,N\}$ be the configurations of original system, $\mathcal{C}=\{1,2,\dots,n\}$ those of the coarser system, $X$ a deterministic left-stochastic $n\times N$ matrix that maps $\mathcal{C}_{O}$ to $\mathcal{C}$. Consider a left-stochastic $N\times N$ matrix $\Gamma_{O}(t)$ describing the evolution of a given system in terms of the configurations $\mathcal{C}_{O}$ and its coarse grained $n\times n$ version $\Gamma(t)$.

Suppose that (i) $\Gamma_{O}(t)$ is not a dilation of $\Gamma(t)$ and (ii) $\Gamma_{O}(t)$ is divisible at time $t^\prime$, meaning that \eqref{divisibility} applies and there is a left-stochastic $N\times N$ matrix $\Gamma_O(t \leftarrow t^\prime)$ such that
\begin{equation}
\Gamma_O(t) = \Gamma_O(t \leftarrow t^\prime)\Gamma_O(t^\prime).
\label{divisibility_O}
\end{equation}

As explained in subsection \ref{ssec:dilation_coarse}, $\Gamma_{O}(t)$ and $\Gamma(t)$ are related by
\begin{equation}
\Gamma(t) = X \Gamma_O(t) Y(0)
\label{coarse_dynamics1}
\end{equation}
where the $N\times n$ stochastic matrix $Y(t)$ is a right-inverse of $X$. As the procedure holds for every $t$,
\begin{equation}
\Gamma(t^\prime) = X\Gamma_O(t^\prime)Y(0).
\label{coarse_dynamics2}
\end{equation}
Substitute \eqref{coarse_dynamics1} into \eqref{divisibility_O}.
\begin{equation}
\Gamma(t) = X \Gamma_O(t) Y(0) = X \Gamma_O(t \leftarrow t^\prime)\Gamma_O(t^\prime)Y(0).
\label{divisibility_coarse1}
\end{equation}

Inserting a product $\hat Y X$ between $\Gamma_O(t \leftarrow t^\prime)$ and $\Gamma_O(t^\prime)$ such that $\hat Y X = \mathds{1}_{N}$ would transform \eqref{divisibility_coarse1} in \eqref{divisibility} upon use of \eqref{coarse_dynamics1} and \eqref{coarse_dynamics2} but such matrix $\hat Y$ does not exist\footnote{The left-stochastic, deterministic $N\times N_O$ matrix $X$ represents a proper coarse graining, meaning that there is at least one configuration in $\mathcal{C}$ that corresponds to two or more configurations of $\mathcal{C}_O$ grouped together. This implies that $X$ has at least two identical columns and is not of full column rank. Since the existence of a left-inverse of $X$ requires $X$ to be of full column rank, we conclude that $X$ has no left-inverse.}.

By assumption, $\Gamma_O(t)$ is does not correspond to an evolution given by a dilation. We are left with only one possibility for \eqref{divisibility_O} to imply divisibility of the coarse-grained system, equation \eqref{divisibility}: the matrices $\Gamma_O(t \leftarrow t^\prime)$ and $\Gamma_O(t^\prime)$ satisfy:
\begin{equation}
X \Gamma_O(t \leftarrow t^\prime)\Gamma_O(t^\prime)Y(0)
=
X \Gamma_O(t \leftarrow t^\prime)Y(t^\prime)X\Gamma_O(t^\prime)Y(0)
\label{divisibility_coarse2}
\end{equation}
for some choice $N\times n$ stochastic $Y(t^\prime)$ such that $XY(t^\prime)=\mathds{1}_n$.

We supposed that $\Gamma_O(t)$ was not itself a dilation but no such assumption was made about $\Gamma_O(t \leftarrow t^\prime)$. It turns out that if $\Gamma_O(t \leftarrow t^\prime)$ is a dilation by coarse graining, proper or under uncertainty, from $\mathcal{C}$ to $\mathcal{C}_O$ then, \eqref{divisibility_coarse2} is trivially solved.
Explicitly,
\begin{align}
\Gamma_O(t) = \left[\sum_{k=0}^{nN}\alpha_k Y_k \Gamma(t \leftarrow t^\prime) X_k\right]\Gamma_O(t^\prime)
\label{divisible_mixed}
\end{align}
implies
\begin{align}
\Gamma(t)
&= X \Gamma_O(t) Y(0)\nonumber\\
&= \Gamma(t \leftarrow t^\prime)\left[X\Gamma_O(t^\prime)Y(0)\right]\nonumber\\
&= \Gamma(t \leftarrow t^\prime)\Gamma(t^\prime).
\label{divisibility_O_degenerate}
\end{align}
where we made use of equations \eqref{coarse_dynamics1} and \eqref{coarse_dynamics2}.
One consequence is that part of the degenerate matrices in any dimension will be divisible according to \eqref{divisible_mixed} which is a higher-dimensional generalisation of \eqref{non_invertible}. One should be aware that, in general, the convex combination of degenerate matrices is not necessarily degenerate.

The most general solution to \eqref{divisibility_coarse2} is yet to be determined.

As an example, consider a system with $\mathcal{C}_O=\{A,B,C\}$ with dynamics for given by
\begin{equation}
\Gamma_O(t)
=
\begin{pmatrix}
1 & 0 & 0\\
0 & \cos^2(\omega t) & \sin^2(\omega t)\\
0 & \sin^2(\omega t) & \cos^2(\omega t)
\end{pmatrix}
\label{cos3}
\end{equation}
The curve given by \eqref{cos3} in the space of stochastic matrices corresponds to an indivisible evolution\footnote{The analysis of the previous section applies here since any $\Gamma_O(t\leftarrow t^\prime)$ that satisfies \eqref{divisibility_O} for \eqref{cos3} is also of the form $\begin{pmatrix}1& \vec{0}^T \\ \vec{0}& \Gamma_{2\times 2}(t\leftarrow t^\prime) \end{pmatrix}$}.

Our goal is to reduce the dynamics for  $\mathcal{C}_O=\{A,B,C\}$ to the dynamics for $\mathcal{C}=\{1, 2\}$. For this, there are different available choices of coarse graining matrix $X$. We start with one that maps an indivisible evolution into an indivisible evolution while extending the duration of the divisible period.
\begin{equation}
X_1 = \begin{pmatrix}1 & 1 & 0 \\ 0 & 0 & 1\end{pmatrix}.
\end{equation}
Its right-inverses form a one parameter family of matrices
\begin{equation}
Y_1\left(p_{Y_1}^{}\right) = \begin{pmatrix} p_{Y_1}^{} & 0 \\ 1-p_{Y_1}^{} & 0 \\ 0 & 1\end{pmatrix}.
\end{equation}
with $p_{Y_1}^{}\geqslant 0$ ensuring all the $Y_1$ are left-stochastic.
From the previous procedure, the reduced matrix associated with $X_1$ is
\begin{equation}
\Gamma_{R_1}^{}(t)
=
\begin{pmatrix}
p_{Y_1}^{}+\left(1-p_{Y_1}^{}\right) \cos^2(\omega t) & \sin^2(\omega t)\\
\left(1-p_{Y_1}^{}\right) \sin^2(\omega t) & \cos^2(\omega t)
\end{pmatrix}
\label{cos3_R1}
\end{equation}
which represents different kinds of indivisible evolution depending on the value of $p_{Y_1}^{}$. Any choice of $p_{Y_1}^{}\neq 0$ leads to an evolution which is not unistochastic in the reduced space: an oscillation between the identity matrix and a matrix at the lower border of the square of figure \ref{fig:02}. The choice of a uniform coarse graining, $p_{Y_1}^{}=1/2$, selects the midpoint of that segment as the turning point of the oscillator. Starting at configuration 1, the chance of finding the system at configuration $2$ is bounded by $p_{Y_1}^{}$.

By applying the reasoning illustrated in figures \ref{fig:09_10} and \ref{fig:17}, we can deduce that the evolution will be divisible for
\begin{equation}
t\in\left[0,\frac{1}{\omega}\cos^{-1}\sqrt{\frac{1-p_{Y_1}^{}}{2-p_{Y_1}^{}}}\right]
\end{equation}
which corresponds to a straight line from the identity matrix until the secondary diagonal with a duration that decreases monotonically with $p_{Y_1}^{}$.

A second option is to map configurations $B$ and $C$ in $\mathcal{C}_O$ to configuration $2$ in $\mathcal{C}$:
\begin{equation}
X_2 = \begin{pmatrix}1 & 0 & 0 \\ 0 & 1 & 1\end{pmatrix}.
\end{equation}
Its right inverses also form a one-parameter family of matrices
\begin{equation}
Y_2\left(p_{Y_2}^{}\right) = \begin{pmatrix} 1 & 0\\0 & p_{Y_2}^{} \\ 0 &1-p_{Y_2}^{}\end{pmatrix}.
\end{equation}
where $p_{Y_2}^{}\geqslant0$ ensures the left-stochasticity of ${Y_2}$. Any choice $p_{Y_2}^{}$ will send any stochastic matrices, divisible or not, of the form
\begin{equation}
\begin{pmatrix}
1 & 0 & 0\\
0 & \Gamma_{22}(t) & 1-\Gamma_{33}(t)\\
0 & 1-\Gamma_{22}(t) & \Gamma_{33}(t)
\end{pmatrix}
\label{gamma_block}
\end{equation}
to the identity in $\mathcal{C}$:
\begin{equation}
\Gamma_{R_2}^{}(t)=
\begin{pmatrix}
1 & 0\\
0 & 1
\end{pmatrix}
\end{equation}

We conclude that the divisibility of the dynamics of a coarse grained system does not imply the divisibility of the original system and there will be cases where \eqref{divisibility_coarse2} will have no solution. An open question is whether it is possible or not that a divisible dynamics reduces to an indivisible one after proper coarse graining\footnote{Although strange at first glance, recall that \eqref{divisibility_O} does not entail \eqref{divisibility} with a reduced $\Gamma(t\leftarrow t^\prime)$ from $\Gamma_O(t\leftarrow t^\prime)$ when \eqref{divisibility_coarse2} is not satisfied for some $\hat Y$. The obstruction to such a reduction by coarse graining does not prevent, nor does it seem to guarantee, the existence of a matrix $\Gamma(t\leftarrow t^\prime)$, {\it not} reduced from $\Gamma_O(t\leftarrow t^\prime)$, satisfying \eqref{divisibility}.}. When coarse graining transforms an indivisible dynamics to an indivisible one, another open question is if the duration of the periods of indivisible evolution must, in general, decrease under coarse graining, as happened in the example with the choice of $X_1$.

\subsection{Ensembles}
\label{ssec:ensembles}

Negligible interactions and absence of entanglement are one way of justifying the treatment of a system as a sum of its parts. Ensembles of systems with indivisible dynamics appear natural in this context, as when each of the items of the ensemble follows, for example, the dynamical law \ref{Gamma_cos_t} or variations of it with different frequencies. In some cases, the interaction may be non-negligible and each part may still be treated separately. In this same oscillator example, the frequency might be associated with quantum spins in an external magnetic field.

Which interactions to neglect are, of course, a matter of convenience as one may use the present approach to consider ensembles of 2-level systems with the well-known decay process modelled by
\begin{equation}
\Gamma(t)=
\begin{pmatrix}
1 & 1-e^{-\lambda t}\\
0 & e^{-\lambda t}
\end{pmatrix}, \quad t\in\mathbb{R}_+.
\label{decay}
\end{equation}
an infinitely divisible case. In our representation in the square, $\Gamma(t)$ is a curve starting at the identity at $(1,1)$ and moving towards $(1,0)$ along the right edge of the square, without ever arriving at the projector $\Pi_1$ at $(1,0)$. A large number of independent copies will provide a system under a homogeneous Poisson process, once all copies start at configuration 2 and one counts the probability of finding $k$ of them at configuration 1.

A more ``interacting" application of the stochastic approach presented here is to 2-level systems in the presence of some symmetry such as spins in a solid state lattice. Each spin is a 2-level system describable in the square representation. Every spin related to another by symmetry, as lattice translations, will be subject to the same dynamical laws except, and it seems very likely, for not necessarily sharing the same division events, a condition which seems even more likely to be true when considering pairs farther and farther apart in space and with the system away from a phase transition. Modelling such systems using {\it indivisible stochastic} dynamics only could provide new perspectives the possible time-evolutions and characterise an avenue for a connection with quantum walks \cite{Venegas-Andraca:2012zkr}. This problem is intimately related to the description of subsystems given an overall indivisible dynamics which would encodes, among other things, the information about the maximum indivisibility and proper division events of its parts, motivating an development of the subsystem analysis here performed.

\section{Conclusions and outlook}
\label{sec:conclusions_and_outlook}

The present work explored the (in)divisibility of stochastic dynamics, focusing on 2 configurations and relating them to larger systems, through symmetries, coarse-graining and dilations. We introduced a geometrical characterisation of divisibility of the time-evolution that shows the existence of a cone structure in the space of $2\times 2$ stochastic matrices separating divisible and indivisible evolutions. We used and extended the work of Schmidt \cite{Schmidt:2021} on dilations by coarse graining in two ways: by introducing a dynamical element and by allowing for uncertainty in the grouping of configurations. Both were shown to play a role in identifying cases when the coarse-graining of a divisible dynamics is also divisible. Although every dilation of a divisible map is divisible, the problem translating the (in)divisibility of a large system to its coarse-grained version is far from solved. Its form strongly suggests the necessity of characterising indivisible dynamics in the space of stochastic matrices in a clearer and more geometrical way, motivating the continuation of this study for $N\geqslant3$.

Bausch and Cubitt showed that finding a stochastic matrix which is the square-root of another stochastic matrix can be solved in polynomial time \cite{Bausch:2016}. They refer as divisible to a stochastic matrix with a stochastic square-root, a notion which differs from ours as we follow reference \cite{Barandes:2023ivl,Barandes:2023pwy}, an approach which, for stochastic dynamics, has the same spirit of \cite{Wolf:2008kbg,Plenio:2014nyj,LiLi:2018ysq, Chruscinski:2022hvy,Ende:2024ksq, Nery:2024fyl}, but operating in a smaller spaces and with real variables. The goal of understanding divisibility from a fundamental perspective places a geometric understanding as worth pursuing, regardless of its impact in computational complexity since, as shown here, it may involve connections between different areas of physics.

The insistence on divisibility and coarse-graining is justified by the goal of characterising indivisibility across scales through different renormalisation methods, providing anchors for treating systems with infinitely many configurations on safer grounds as these undergo  indivisible dynamics, with the hope of a greater ease to apply the theoretical framework to experimentally data. The path proposed here is to avoid, as much as it is possible and fruitful, the help of complex Hilbert spaces in order to extract the maximum that indivisibility may supply directly from stochastic dynamics. In this direction, an unexplored direction consists in exploring the subgraphs of the stochastic matrix graph along the time evolution, in particular using the framework of Hopf monoids which provide a structure to handle both permutahedra and subgraphs, with different operations of joining and dividing graphs \cite{Aguiar:2017}, which seem physically clear and could be relevant to understand higher order spaces of stochastic matrices, the stochastic polytopes.

\section*{Acknowledgments}

The author would like to thank Lucas Tavares Cardoso, Rafael Chaves Souto Ara\'ujo, Dmitry Melnikov, Rodrigo Pereira, Daniel Augusto Turolla Vanzella, Diogo de Oliveira Soares-Pinto for helpful conversations and specially for D. O. Soares-Pinto and L. T. Cardoso for comments on different versions of this manuscript. The author also thanks Analia Silva, Sebasti\~ao das Gra\c{c}as Pimenta and Laura Keil for support during important steps of this project.

\bibliographystyle{apsrev4-2}

\end{document}